\documentclass[twocolumn]{aastex631}
\usepackage{natbib,graphics,graphicx,multirow,amsmath,cancel,color,xcolor,hyperref}
\usepackage{soul}
\usepackage[normalem]{ulem}
\usepackage{float}
\usepackage{comment}
\usepackage{tabularx}
\usepackage{multirow}
\usepackage{array}
\renewcommand{\arraystretch}{1.2}
\usepackage{color, colortbl}

\begin{document}
\title{Contribution of spicules to solar coronal emission}

\correspondingauthor{Shanwlee Sow Mondal}
\email{shanwlee@prl.res.in\\
shanwlee.sowmondal@gmail.com}

\author[0000-0003-4225-8520]{Shanwlee Sow Mondal}
\affil{Astronomy and Astrophysics Division, Physical Research Laboratory, Ahmedabad 380009, India}
\affil{Indian Institute of Technology, Gandhinagar, Gujarat 382355, India}

\author[0000-0003-2255-0305]{James A. Klimchuk}
\affil{Heliophysics Science Division, NASA Goddard Space Flight Center, 8800 Greenbelt Rd., Greenbelt, MD 20771, USA}

\author[0000-0002-4781-5798]{Aveek Sarkar}
\affil{Astronomy and Astrophysics Division, Physical Research Laboratory, Ahmedabad 380009, India}

\begin{abstract}
Recent high-resolution imaging and spectroscopic observations have generated renewed interest in spicules' role in explaining the hot corona. Some studies suggest that some spicules, often classified as type II, may provide significant mass and energy to the corona. Here we use numerical simulations to investigate whether such spicules can produce the observed coronal emission without any additional coronal heating agent. Model spicules consisting of a cold body and hot tip are injected into the base of a warm ($0.5$ MK) equilibrium loop with different tip temperatures and injection velocities. Both piston- and pressure-driven shocks are produced. We find that the hot tip cools rapidly and disappears from coronal emission lines such as Fe XII $195$ and Fe XIV $274$. Prolonged hot emission is produced by pre-existing loop material heated by the shock and by thermal conduction from the shock. However, the shapes and Doppler shifts of synthetic line profiles show significant discrepancies with observations.
Furthermore, spatially and temporally averaged intensities are extremely low, suggesting that if the observed intensities from the quiet Sun and active regions were solely due to type II spicules, one to several orders of magnitude more spicules would be required than have been reported in the literature. This conclusion applies strictly to the ejected spicular material. We make no claims about emissions connected with waves or coronal currents that may be generated during the ejection process and heat the surrounding area. 
\end{abstract}
\section{Introduction}\label{sec:introduction}
Defying decades of continued efforts, many aspects of coronal heating remain unanswered~\citep{Klimchuk_2006, Klimchuk_2015, Viall_2021}. Even the basic mechanism is a matter of debate. Despite the fact that all the coronal mass is sourced at the chromosphere, agreement on how the chromospheric mass is heated and transported up to the corona has not been reached. An early observation of the solar chromosphere reported the existence of several small jet-like features~\citep{Secchi1877}. They were later named~\textit{spicules}~\citep{Roberts45}. With improved observations, these spicules were seen to propagate upwards~\citep{Pneuman_1977, Pneuman_1978} with speed $20$ - $50$ km s$^{-1}$. They were also seen to survive for about $5$ to $10$ minutes and carry almost $100$ times the mass needed to balance the mass loss in the solar corona due to the solar wind. Further studies of the spicules~\citep{Athay_1982} suggested a pivotal role in transferring energy from the inner layers of the solar atmosphere to the lower solar corona. However, the proposal was not pursued further because these traditional spicules lack emission in the Transition Region (TR) and coronal lines~\citep{Withbroe_1983}.

About a decade ago, using high-resolution imaging and spectroscopic observations from the Hinode and Solar Dynamic Observatory missions, \cite{Pontieu_2007, Pontieu_2011} further discovered jet like features traveling from the chromosphere to the corona. These features appear all over the Sun with a lifetime between $10-150$ s and a velocity of $50-150$ km s$^{-1}$. \cite{Pontieu_2007} termed them type II spicules and suggested that they are capable of connecting the relatively cooler solar chromosphere with the hot corona.  

Since their discovery, multiple observations have identified type II spicules and reported on their characteristics. However, nothing conclusive has yet been established about their origin. Only recently,~\cite{samanta_19} have identified the near-simultaneous origin of spicules and the emergence of photospheric magnetic bipoles. The tips of the originated spicules eventually appear in the coronal passband, suggesting that the plasma is heated to coronal temperatures. A 2.5D radiative MHD simulation of type II spicules~\citep{Martinez_2017} has reproduced many of their observed features. This simulation also suggests that ambipolar diffusion in the partially ionized chromosphere may play a crucial role in the origin of type II spicules. On the other hand, a recent work~\citep{Dey_22} based on radiative MHD simulation and laboratory experiment suggests that quasi-periodic photospheric driving in the presence of vertical magnetic fields can readily generate spicules in the solar atmosphere. Their work, devoid of any chromospheric physics, can still account for the abundance of wide varieties of spicules, as seen in the observations.

The evolution of spicules during their propagation is understood through multi-wavelength studies~\citep[e.g.,][]{Pontieu_2011, Skogsrud_2015}. Observations of ~\cite{Pontieu_2011} suggest that spicule plasma emanating from the chromosphere gets heated to transition region (TR) temperatures and even up to coronal temperatures. Such heating may happen for two reasons:
    \begin{enumerate}
    \item[(a)] Spicule propagation can produce shocks, compressing the material lying ahead of it. In such a scenario, it is not the ejected spicule material but the pre-existing coronal material in front of it that gets compressed by the shock to contribute to the hot emission ~\citep{Klimchuk_2012, Petralia_2014};
    \item[(b)] The tip of the spicule may get heated during the ejection process, on-site, through impulsive heating and produce emissions in the coronal lines. In the latter scenario, the emission indeed comes from the ejected spicule material ~\citep{Pontieu_2007}.
    \end{enumerate}
The radiative MHD simulations of ~\cite{Martinez_2018} suggest that spicules and surrounding areas get heated by ohmic dissipation of newly created currents and by waves. Note, however, that the currents in the simulations are relatively large-scale volume currents and would not be dissipated efficiently at the many orders of magnitude smaller resistivity of the real corona. Heating in the real corona involves magnetic reconnection at thin current sheets, of which there are at least $100,000$ in a single active region \citep{Klimchuk_2015}. It is not known whether the ohmic heating in the simulations is a good proxy for the actual reconnection-based heating.

\cite{Klimchuk_2012} considered a simple analytical model for the evolution of spicules with a hot tip. He argued that if a majority of observed coronal emission were from such hot tips, it would be inconsistent with several observational features (see also \cite{Tripathi_2013, Patsourakos_2014}). The result was further supported by hydrodynamic simulations~\citep{Klimchuk_2014}. Using these simulations, they studied the response of a static loop to impulsive heating in the upper chromosphere, which produces localized hot material that rapidly expands upward and might represent the hot tip of a spicule. Noticing the inability of a single hot spicule tip to explain the observations, \cite{Bradshaw_2015} further explored the role of frequently recurring chromospheric nanoflares. The study was motivated by the suggestion that rapidly repeating type II spicules might accumulate enough hot plasma to explain the coronal observations  \citep{Pontieu_2011}. However, the simulations were still inconsistent with observations. 

In both the analytical model and the simulations, the dynamics of the hot material is due entirely to an explosive expansion from the locally enhanced pressure. There is no additional imposed force to bodily eject the material. The consequences of such a force were investigated by
 \cite{Petralia_2014}. Their study involves injecting cold and dense chromospheric material into the corona with an initial velocity. The result indicates that the production of a shock can give rise to coronal emission. However, the emission is from the preexisting coronal material rather than the spicule itself. The injected material has no hot component. 

The studies mentioned above have investigated the dynamics of either the hot tip of a spicule without any initial velocity or a spicule with a cold tip and finite injection velocity. Our work combines these two effects. The spicule is now injected in a stratified flux tube with high velocity and consists of both a hot tip and a cold body ($T = 2 \times 10^{4}$~K). We further investigate the possibility that most of the observed hot emission from the corona can be explained by such spicules. Through forward modelling, we quantitatively compare the simulations with observations to answer this question.

The rest of this paper is organized as follows. The numerical setup is described in Section~\ref{sec:setup}. We report on the simulation results in Section~\ref{sec:result}. Finally we summarize and discuss our results in Section~\ref{sec:summary}.

\section{Numerical Setup}\label{sec:setup}
Spicules are seen to follow magnetic field lines. To simulate their dynamics, we consider a straight 2D magnetic flux tube consisting of uniform $10$~G magnetic field. We impose a gravity corresponding to a semi-circular loop  such that the vertical component of the gravitational force is maximum at both ends and smoothly becomes zero in the middle of the tube. Two ends of the tube are embedded in the chromosphere. The loop is symmetric about the center, which corresponds to the apex. We use Cartesian coordinates, and therefore the loop actually corresponds to an infinite slab. This is a reasonable approximation because we are interested in how the plasma evolves within an effectively rigid magnetic field appropriate to the low $\beta$ corona. The slab dimension corresponding to the loop length is $100$~Mm. The other dimension is $0.42$~Mm, but this is not relevant. Rigid wall boundary conditions are imposed at the sides, and the evolution is essentially equivalent to 1D hydrodynamics, as discussed below. The first 2 Mm of both ends of the loop are resolved with a fine uniform grid with 10 km cells, while the coronal part is resolved with a stretched grid containing 1500 cells on each side. The fine grid close to both the footpoints allows us to resolve the steep transition region more accurately. 

\begin{figure}
\centering
	 \includegraphics[width=0.48 \textwidth, angle=0]{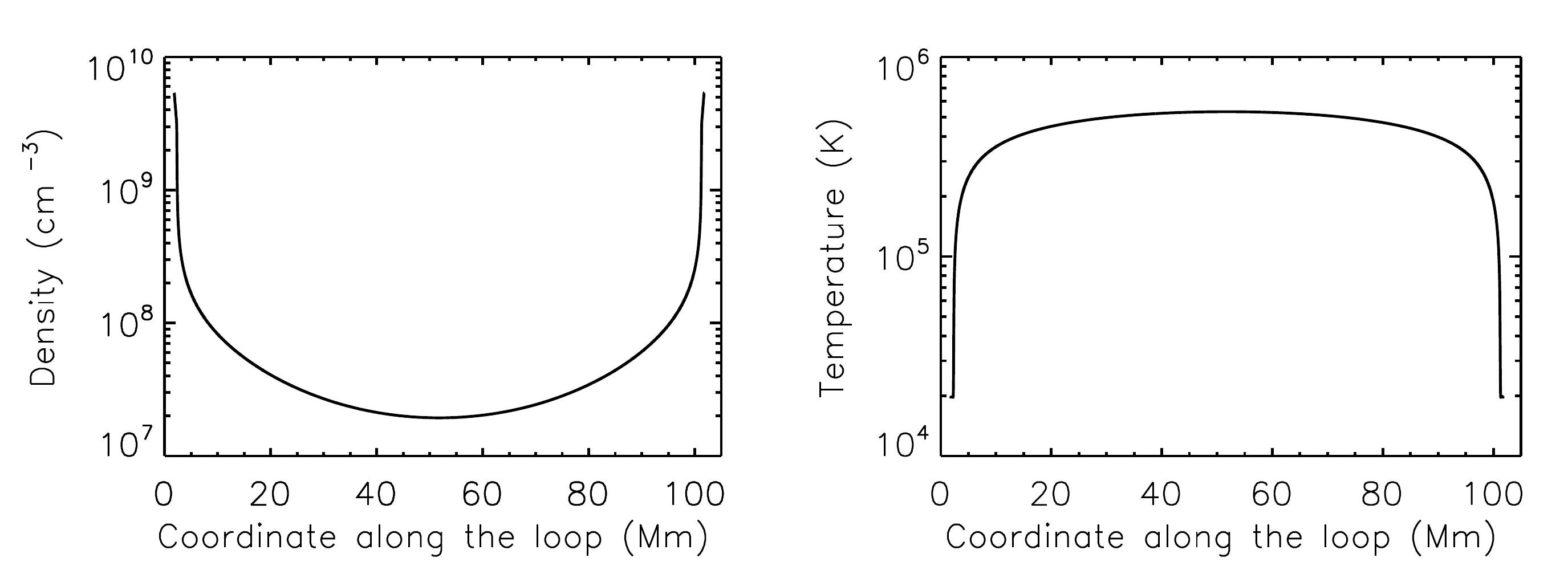}
	 \caption{Density and temperature profiles of the initial static equilibrium loop.}
	 \label{initial_density_temp}
\end{figure}

The spicule simulation begins with an initial static equilibrium atmosphere obtained with the double relaxation method described in Appendix~\ref{append:steady_state}. We choose a relatively low temperature and low density loop because we wish to test the hypothesis that the observed coronal emission comes primarily from spicules. The apex temperature of the loop is $0.5$ MK. Figure~\ref{initial_density_temp} shows the background loop profile that is used in most of our simulations. The chromosphere is $470$ km deep - approximately half a gravitational scale height. It merely acts as a mass reservoir. Detailed chromospheric physics like partial ionization and optically thick radiation are not implemented in the code as we are solely interested in coronal emission. We use a modified radiation loss function to maintain a chromospheric temperature near $2 \times 10^{4}$~K, as described in Appendix~\ref{append:steady_state}.

The propagation of a spicule in the loop is emulated through an injection of dense material from the left footpoint. The injected material follows specified density, velocity and temperature profiles in time which are described below. At this injection boundary, all plasma parameters, except the density and pressure, are set to their initial values once the injection is over. The density is set to have the prescribed value at the end of the injection phase, and pressure is determined from the ideal gas equation of state. On the other hand, at the right footpoint, all the plasma parameters maintain the initially prescribed values throughout the entire simulation. 

We solve the compressible MHD equations inside our simulation domain using the PLUTO code \citep{2007ApJS..170..228M} with ideal gas environment. Plasma inside the domain is cooled by radiation and field aligned thermal conduction. The CHIANTI \citep{chianti} radiative loss function for coronal abundance is used to model the radiative cooling. For anisotropic conduction, the thermal conductivity, $\kappa_{\parallel}$ = $5.6 \times 10^{-7} T^{5/2}$ erg s$^{-1}$ K$^{-1}$ cm$^{-1}$ is considered along the magnetic field lines, whereas $\kappa_{\perp}$ is taken to be zero. Also, for the saturated conductive flux used in PLUTO, $F_{sat}$ = 5$\phi \rho C_{s}^{3}$, where we have considered the value of the free parameter $\phi$ to be 0.9, which represents effective thermal conduction in the system, and $C_{s}$ is the isothermal sound speed. The MHD equations are solved in Cartesian coordinates. 

The photospheric magnetic flux is found to be localized and clumpy, whereas, in the corona, it fills out space uniformly. Such nature of the magnetic flux at different layers of the solar atmosphere dictates that the flux tubes expand laterally at the junction of the chromosphere and corona, where the plasma $\beta$ changes from being greater than one to less than one. This type of expansion of flux tubes is realized in 2D MHD simulations of coronal loops~\citep[e.g.,][]{Guarrasi_14}. Through an area expansion factor, this has also been incorporated in 1D or 0D models \citep{Mikic_13, Cargill_22}. We do not include expansion in our model because we are interested in the spicule dynamics in the corona, and the simplification should not affect our results significantly. We note that the plasma $\beta$ is less than unity throughout the evolution, so no expansion from the spicule injection would be expected.  Additionally, the initial atmosphere and injection profile are uniform along the horizontal (cross-field) axis. Hence the plasma remains nearly uniform in the lateral direction, effectively making our simulations similar to 1D hydrodynamic simulations. Nevertheless, we ran all our computations using the 2D MHD set up because of our familiarity with the powerful PLUTO code. The limited number of grid points in the cross-field direction keeps the computational demands relatively low.

Two main components of our simulations are:  (a) a background loop in hydrostatic and energy equilibrium representing a tenuous coronal atmosphere, and (b) the propagation of injected material resembling spicule propagation along the loop. Our experimental spicule consists of a hot dense tip followed by cold dense material injected from the base of the model. Here we investigate how changing the temperature of the hot tip and injection speed can alter the intensities and profiles of the Fe XII (195 \AA) and Fe XIV(274 \AA) coronal spectral lines. 

\begin{figure*}
\centering
	 \includegraphics[width=0.8 \textwidth,angle=0]{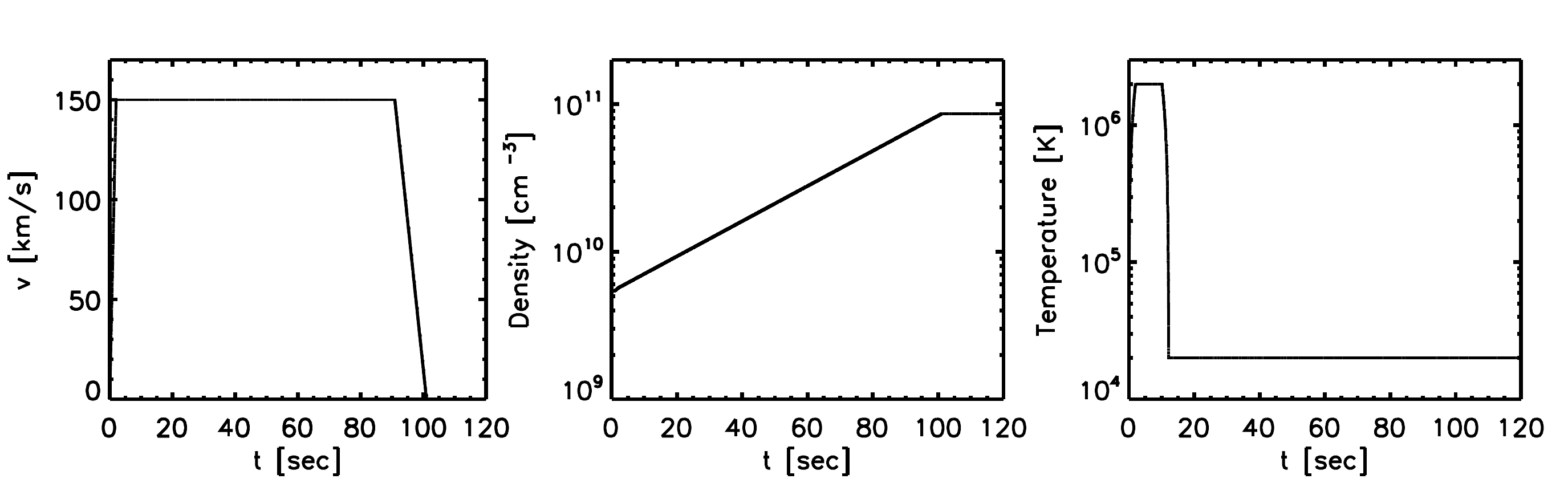}
	 \caption{Injection velocity, density and temperature profiles as a function of time at the bottom boundary from Run1.}
	 \label{pulse}
\end{figure*}

We have performed six sets of simulations where the spicule tip temperatures are considered to be at $2$, $1$, and $0.02$ MK, followed by cold material with a temperature of $0.02$ MK. All the runs are performed with two injection velocities: $50$ and $150$ km s$^{-1}$ (see Table~\ref{tab:table1}). Since we assume that spicules might have been generated deep inside the chromosphere, we inject a high-density material in the loop to emulate the spicule. The density scale height of the spicule is chosen to be six times the gravitational scale height at the base of the equilibrium loop. To impose such conditions on the ejected spicule, its density follows a time profile given by,

\begin{equation}
\label{rho_profile}
  \rho(t)=
  \begin{cases}
    \rho_{0}\exp\Big[\frac{v(t) t}{6H}\Big], & \ 0 < t \le t_{5} \\
    \rho(t_{5}), & \ t_{5} < t \\
      \end{cases}~,
\end{equation}
 where $\rho(t)$ and $\rho_{0}$ are the injected density at time $t$ and the base density of the equilibrium loop, respectively.  The time profile of the injection velocity is given by,
 \begin{equation}
\label{vel_profile}
  v(t)=
  \begin{cases}
    v_{inj} \times \Big(\frac{t}{t_{1}}\Big), & \ 0 < t \le t_{1} \\
    v_{inj}, & \ t_{1} < t \le t_{4}\\
    v_{inj} \times \Big(\frac{t_{5}-t}{t_{5}-t_{4}}\Big), & \ t_{4} < t \le t_{5} \\
    0, & \ t_{5} < t \\
  \end{cases}~,
\end{equation}
where $v_{inj}$ corresponds to $50$ or $150$ km s$^{-1}$ (depending on the simulation).  $H$ represents the gravitational scale height given by
 
 \begin{equation}
   H =  \frac{k_{B}T_{base}}{\mu m_{H} g_{\odot}}~,
 \end{equation}
where $T_{base}=0.02$~MK is the base temperature of the loop, $k_{B}$ is the Boltzmann constant, while $m_{H}$ and $g_{\odot}$ represent mass of the hydrogen atom and solar surface gravity, respectively, and $\mu = 0.67$ denotes the mean molecular weight of the plasma.  
 The temperature of the ejected spicule also follows a time profile given by
\begin{equation}
\label{tmp_profile}
  T(t) =
  \begin{cases}
    T_{base} + (T_{tip}-T_{base})\times\Big(\frac{t}{t_{1}}\Big), & \ 0 < t \le t_{1} \\
    T_{tip}, & \ t_{1} < t \le t_{2}\\
    T_{tip} + (T_{base}-T_{tip})\times\Big(\frac{t-t_{2}}{t_{3}-t_{2}}\Big), & \ t_{2} < t \le t_{3} \\
    T_{base}, & \ t_{3} < t \\
  \end{cases}~,
\end{equation}
where $T_{base}$ is the temperature of the cold material (bottom part) of the spicule ($=0.02$ MK) and  $T_{tip}$ is the spicule tip temperature which can take values $2$, $1$, or $0.02$~MK depending on the run being performed. In all the above equations, times $t_{1}$, $t_{2}$, $t_{3}$, $t_{4}$ and $t_{5}$ are chosen to be $2$, $10$, $12$, $90$ and $100$ s, respectively. Times are chosen so that the top $10\%$ of the spicule's body emits in coronal lines as is generally observed~\citep{Pontieu_2011}. The total injection duration is also motivated by the observed lifetime of type II spicules \citep{Pontieu_2011}.
The ramping up of velocity, density, and temperature ensures a smooth entry of the spicules into the simulation domain. Similarly, the ramping down at the end of the injection avoids any spurious effects. Figure~\ref{pulse} shows one such example of velocity, density, and temperature profiles when the spicule is ejected with velocity $150$ km s$^{-1}$, and its hot tip is at $2$~MK. Likewise, different injection time profiles have been used for other injection velocities and temperatures. The initial equilibrium loop remains the same in all cases, unless specified. \\

\section{Results}\label{sec:result}
The large velocity of the spicule and high pressure compared to the ambient medium give rise to a shock, which propagates along the loop and heats the material ahead of it. Depending on the sound speed of the ambient medium (i.e., the preexisting loop plasma) and the temperature of the injected spicule material, the generated shock turns out to be of different kinds: (a) Piston driven shock -- in which case the shock speed is nearly equal to the injection speed (e.g., simulation with $T_{tip} = 0.02$~MK), and (b) Pressure driven~ shock -- in which case the shock speed exceeds the injection speed (e.g., when $T_{tip} = 2$ and $1$~MK). Emission from the shock heated plasma differs depending on the nature of the shock. 

We compare different simulations to understand the coronal response to spicules with different injection parameters. Our discussion starts with the results from Run1 where the hot tip of the injected spicule has a temperature $T_{tip} = 2$~MK and injection velocity $v = 150$ km s$^{-1}$. The injection profiles are those already shown in Figure~\ref{pulse}.

\subsection{Dynamics and heating}
\begin{figure}
\centering
	 \includegraphics[width=0.48 \textwidth,angle=0]{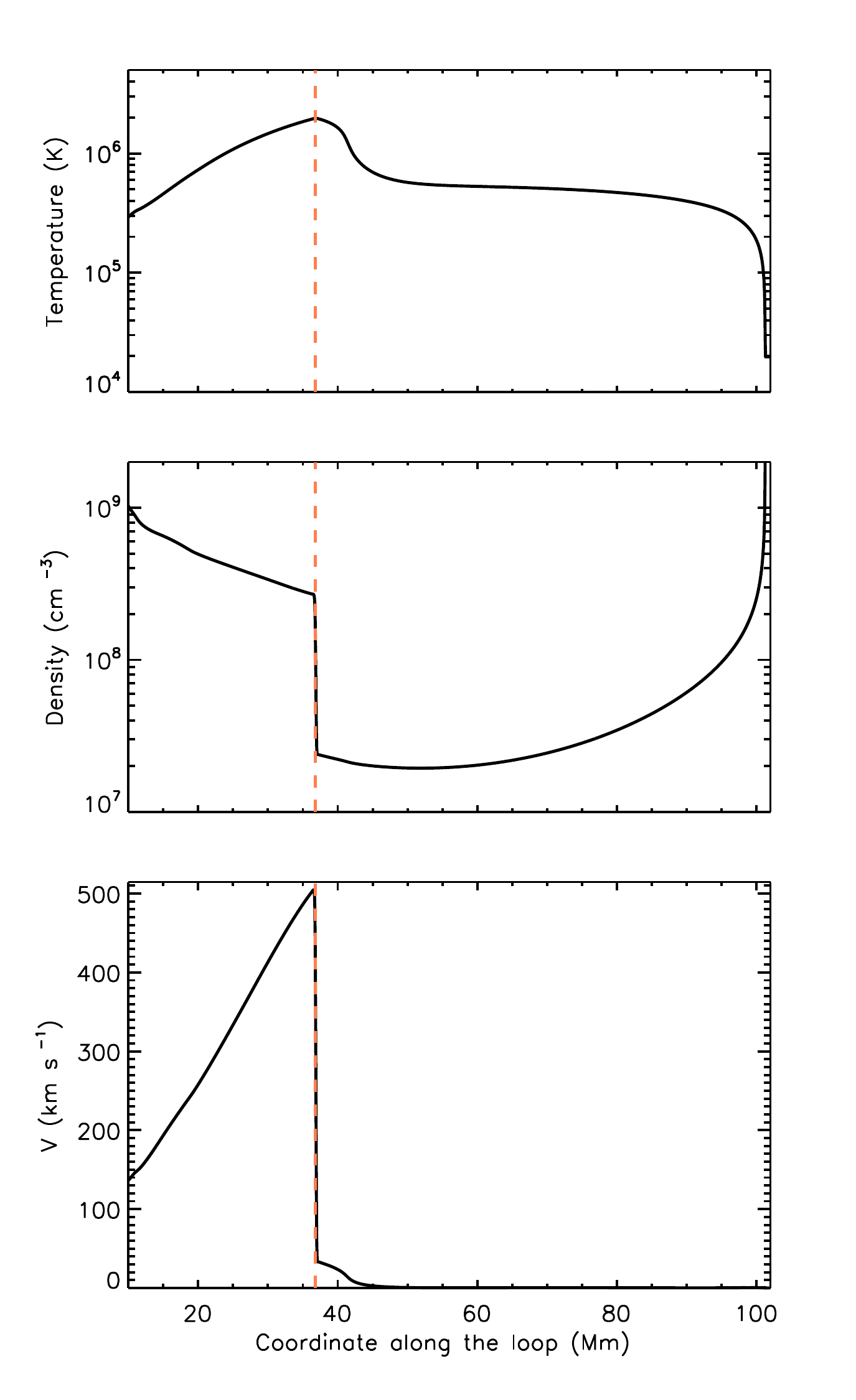}
	 \caption{Plasma variables along the loop at $t = 70$~s in Run1. The colored dashed line marks the location of shock front. The observed density and velocity jumps match well with $\gamma=1.015$ indicating the shock is locally isothermal, consistent with the temperature profile.}
	 \label{shock}
\end{figure}

Insertion of dense, high temperature plasma ($T_{tip} = 2$ MK) with high velocity ($v = 150$ km s$^{-1}$, Run1) into the warm loop produces a shock. Figure~\ref{shock} shows the temperature, density and plasma velocity along the loop at $t = 70$~s. The dashed lines mark the location of the shock front.  It is evident from the figure that the high compression ratio exceeds the ratio of an adiabatic shock. The compression ratio of an adiabatic shock should always be $\leq 4$. To understand the nature of the shock, we perform a shock test with Rankine-Hugoniot (RH) conditions, which read
\begin{equation}\label{RH_rho_vel}
 \frac{\rho_{2}}{\rho_{1}} = \frac{\gamma + 1}{\frac{2}{M^{2}} + (\gamma - 1)} = \frac{v_{1}}{v_{2}}~.
\end{equation}
Here $\rho_{1}$ and $\rho_{2}$ are the pre- and post-shock plasma mass densities respectively, and $v_{1}$ and $v_{2}$ are likewise the pre- and post-shock plasma velocities in the shock rest frame. Furthermore, $\gamma$ is the ratio of the specific heats, $c_{s} = \sqrt{\frac{\gamma P_{1}}{\rho_{1}}}$ is the upstream sound speed, where $P_{1}$ is the upstream pressure, and finally $M=v_{1}/c_{s}$ is the upstream Mach number in the shock reference frame. Injection of high temperature plasma accelerates the shock with a speed much larger than the injection speed of the spicule material giving rise to a pressure driven shock front. 

Figure~\ref{shock} demonstrates an abrupt change in plasma variables at the shock. The shock speed at this instant is $562$ km s$^{-1}$. It also shows that at the discontinuity location ($s = 36.9$~Mm, $s$ being the coordinate along the loop), the density and velocity ratios are $10.7$ and $0.094$, respectively, in the shock rest frame. The inverse relationship of these ratios indicates a constant mass flux across the shock front, in accordance with equation~\eqref{RH_rho_vel}. The Mach number in the shock frame at the same location is $3.37$. With this Mach number, the RH condition gives density and velocity ratios in accordance with those in the simulation ($10.5$ and $0.095$) when $\gamma = 1.015$. In other words, consistency is achieved with this value of $\gamma$. Being close to unity, it implies a nearly isothermal shock. Efficient thermal conduction carries a large heat flux from the shock front to its surroundings, giving rise to the locally smooth, near-isothermal temperature profile in Figure~\ref{shock}. It is worth mentioning here that RH-jump conditions do not consider any heat loss/gain, such as thermal conduction or radiative loss. However, our system includes these sink terms in the energy equation. It is because of these loss functions the shock-jump is larger. Limited thermal conduction would bring the jump condition closer to the adiabatic approximation but would also affect the thermal profile ahead and behind the shock. Our result is consistent with that of~\cite{Petralia_2014}, where the signature of the shocks in front of the spicule has been reported. As we show later, the initially hot material in the spicule tip cools dramatically. Only ambient material heated by the shock is hot enough to produce significant coronal emission. 

Interestingly, the high compression ratio at the shock front depends more on the temperature difference and corresponding pressure difference between the injected and ambient plasma material than on the velocity with which it is injected. Table~\ref{tab:table1} shows a study of how the compression ratio (or the shock strength) varies when the tip of the spicules are at different temperatures and are injected with different velocities. As mentioned earlier, the injection conditions give rise to two different types of shocks. When the injected plasma temperature is high (e.g., spicule tips with temperatures $2$ and $1$~MK), the excess pressure gives rise to a pressure-driven shock. On the other hand, injection of a cold material (tip temperature equal to that at the loop footpoint, i.e., $0.02$~MK) produces a piston-driven shock. Our test runs identify both kinds of shocks. For example, when we inject spicules with a fixed injection velocity of $150$~km s$^{-1}$, but with  different tip temperatures (viz.\,$2$, $1$ and $0.02$~MK), the average shock speed is $520$, $400$ and $210$ km s$^{-1}$, respectively (see Figure~\ref{shock_speed}). The first two shocks are pressure-driven as the average shock speeds exceed the injection speed by a wide margin. The third shock maintains a speed close to the injection speed and can be categorized as a piston-driven shock. The shock speed depends not only on the injected tip temperature, but also on the properties of the ambient material in which it is propagating, which vary along the loop. This is discussed further in Appendix~\ref{append:shock_speed}.

\begin{deluxetable}{cccc}
\tablenum{1}
\tablecaption{Dependence of compression ratio on the injected hot tip temperature and speed.\label{tab:table1}}
\tablehead{
\colhead{Run} & \colhead{\hspace{1cm}$T_{tip}$\hspace{1cm}} & \colhead{\hspace{1cm}$v$\hspace{1cm}} & \colhead{\hspace{1cm}Compression\hspace{1cm}}\\
\colhead{} & \colhead{\hspace{1cm}(MK)\hspace{1cm}} & \colhead{\hspace{1cm}(km s$^{-1}$)\hspace{1cm}} & \colhead{\hspace{1cm}ratio\hspace{1cm}}
}
\startdata
1 & \hspace{1cm} 2 & \hspace{1cm}150 & \hspace{1cm}11.2\\
2 & \hspace{1cm}2 & \hspace{1cm}50 & \hspace{1cm}8.9\\
3 & \hspace{1cm}1 & \hspace{1cm}150 & \hspace{1cm}8.7\\
4 & \hspace{1cm}1 & \hspace{1cm}50 & \hspace{1cm}6.2\\
5 & \hspace{1cm}0.02 & \hspace{1cm}150 & \hspace{1cm}3.6\\
6 & \hspace{1cm}0.02 & \hspace{1cm}50 & \hspace{1cm}1.7\\
\enddata
\end{deluxetable}

\subsection{Loop emission}\label{sec:forward}
Thermally conducted energy from the shock front heats the material lying ahead of it. Therefore, a magnetic flux tube subjected to spicule activity could produce hot emission from newly ejected material at the spicule's hot tip and from pre-existing coronal material in both the pre and post-shock regions. We now examine the contributions from these three different sources. We identify the leading edge of the hot spicule tip by finding the location in the loop where the column mass integrated from the right footpoint equals the initial column mass of the loop. Recall that the spicule is injected from the left footpoint. The spicule compresses the material in the loop, but does not change its column mass. We identify the trailing edge of the hot material in a similar manner, but using the column mass at time $t=10$~s, when the injection of hot material ceases and the injection of cold material begins.

\begin{figure*}
\centering
	 \includegraphics[width=0.48\textwidth,angle=0]{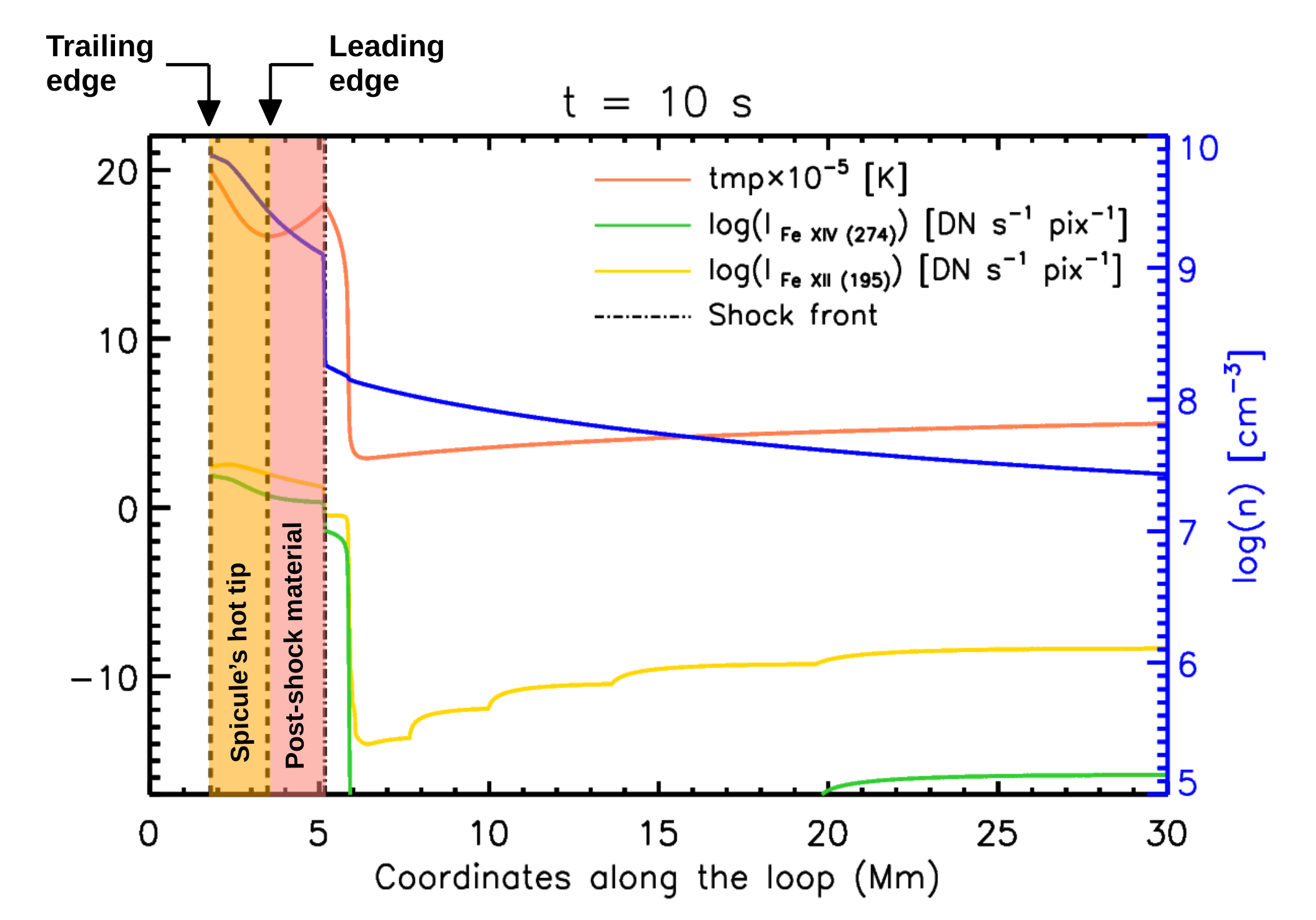}
	 \includegraphics[width=0.48\textwidth,angle=0]{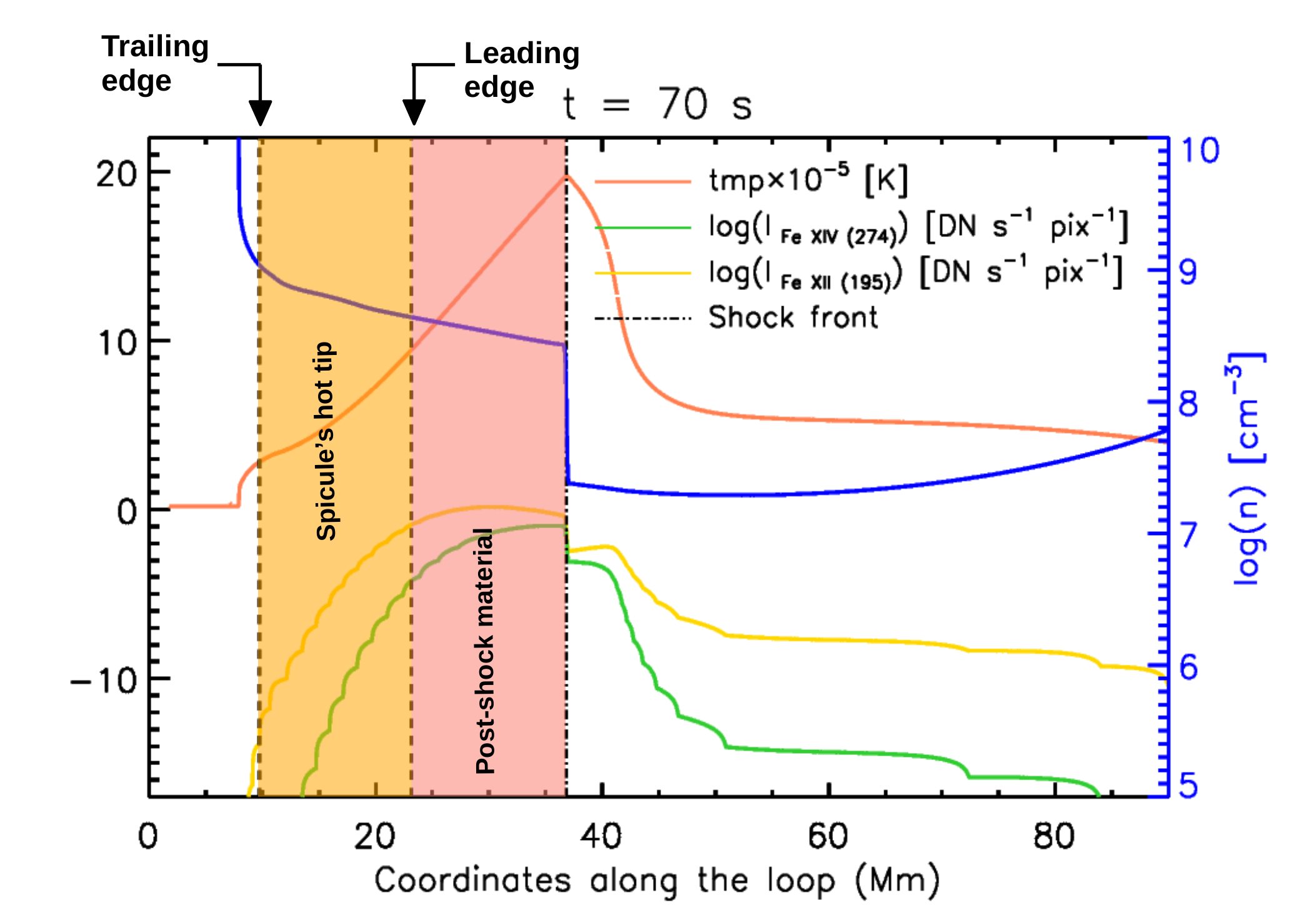}
	 \caption{Logarithm of Fe XII and Fe XIV intensity (yellow and green, respectively) and temperature (dark orange) as a function of position along the loop at t = 10 and 70 s in Run1. Units are given in the legend. The blue curve is the logarithm of density with the scale on the right. The marked light orange region indicates the hot tip of the spicule, while the red region indicates the compressed post-shock plasma. The dot-dashed line (ahead of the red region) marks the location of the shock front.}
	 \label{fe12_10_70}
\end{figure*}

Figure~\ref{fe12_10_70} shows emission along the loop in the Fe XII and Fe XIV lines at $t = 10$ and $70$~s, evaluated from Run1. The orange region is the hot spicule tip, while the red region is the shock-heated material ahead of it. The shock front is the dot-dashed black vertical line. The dark orange curve is temperature in units of $10^5$ K, with the scale on the left. The blue curve is the logarithm of density, with the scale on the right. The yellow and green curves are the logarithms of Fe XII and Fe XIV intensity, respectively, with the scale on the left. The variation of intensity is enormous; a difference of 10 corresponds to 10 orders of magnitude. The intensity is what would be observed by the Extreme ultraviolet Imaging Spectrometer (EIS; \cite{Culhane_2007}) onboard Hinode \citep{Kosugi_2007} if the emitting plasma had a line-of-sight depth equal to the EIS pixel dimension, i.e., if observing an EIS pixel cube. This can be interpreted as normalized emissivity.

\begin{figure*}
\centering
	   \includegraphics[width=0.48 \textwidth,angle=0]{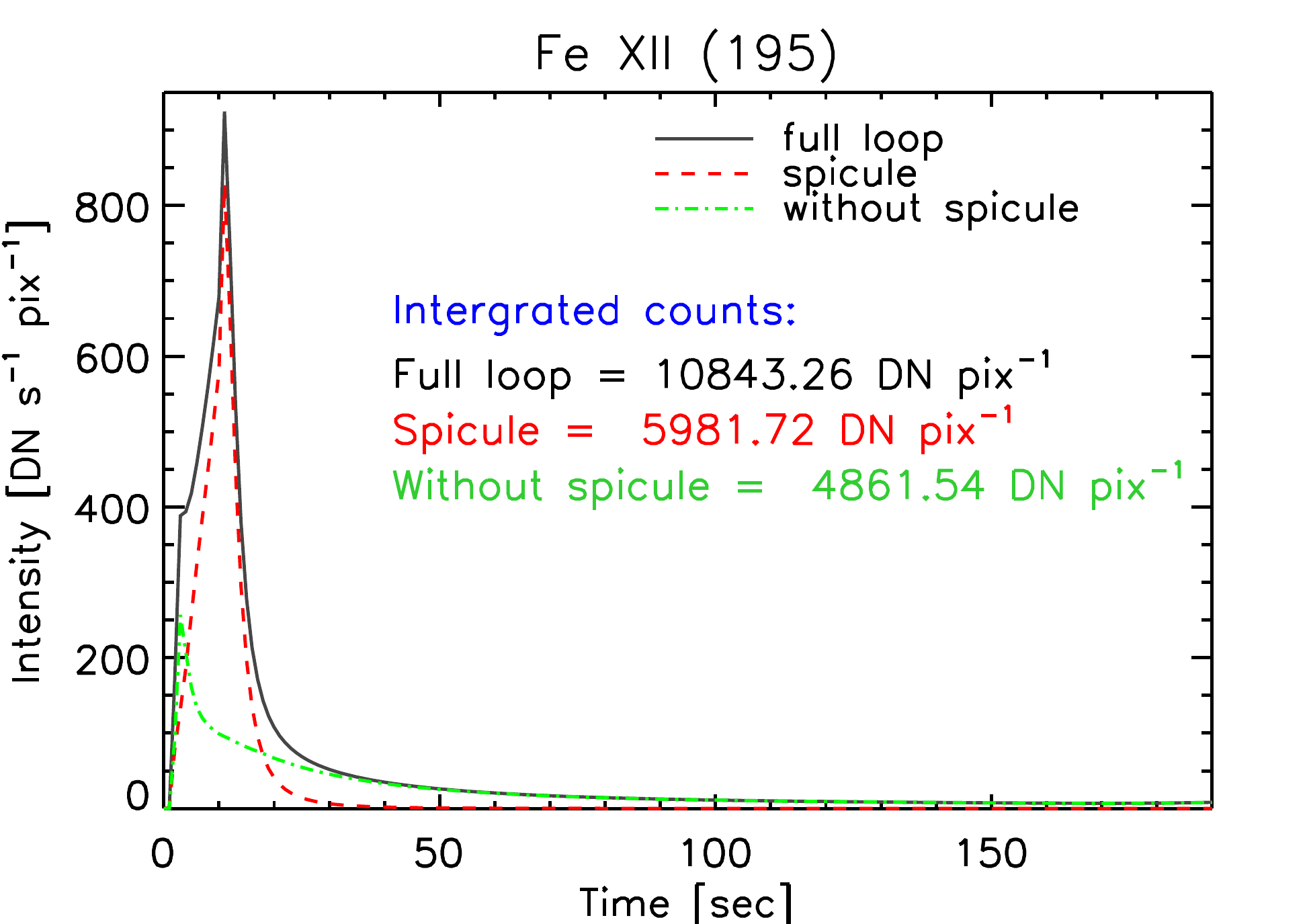}
	  \includegraphics[width=0.48 \textwidth,angle=0]{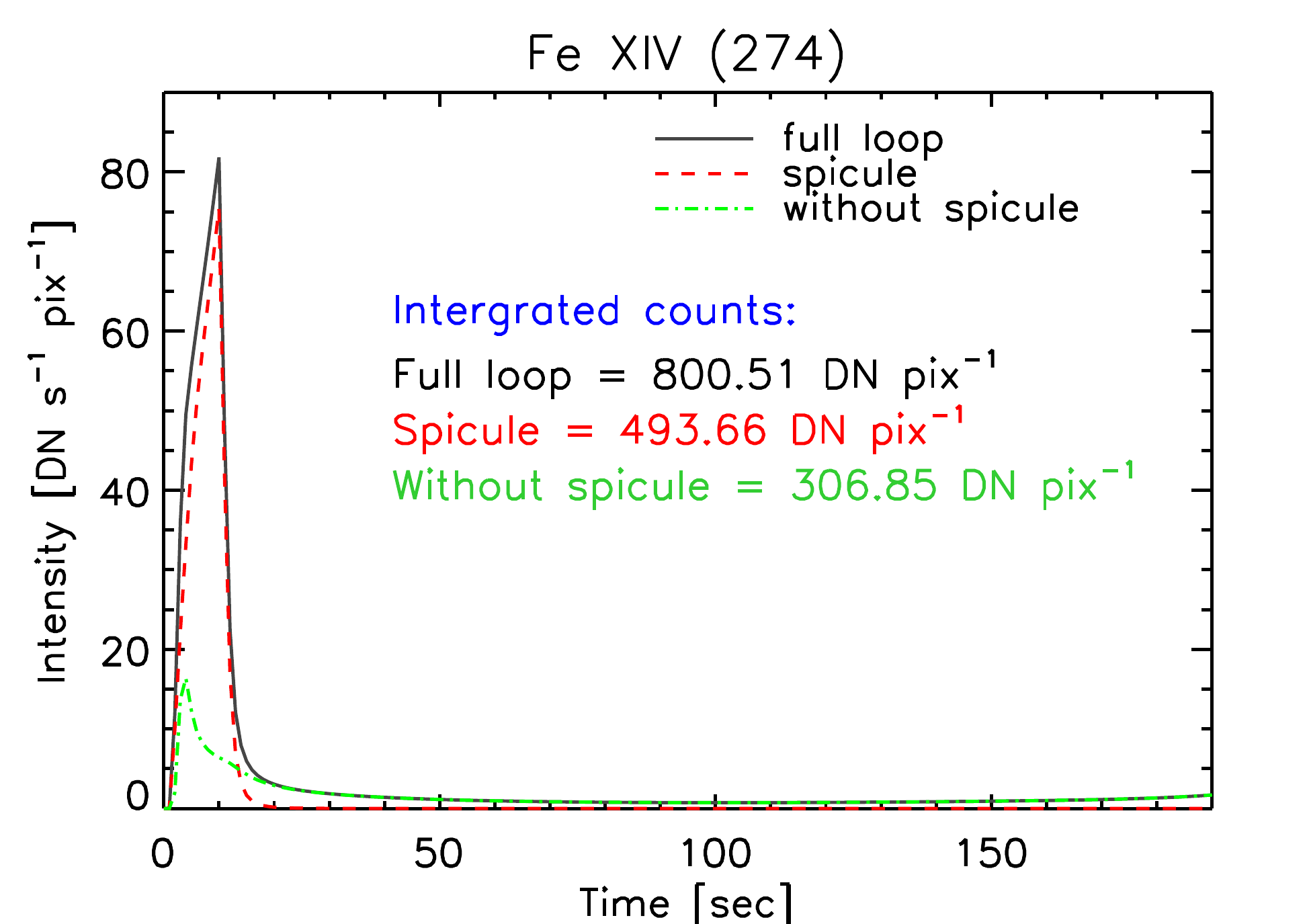}
	 \caption{Loop integrated intensity in Fe XII and Fe XIV as a function of time for Run1. The red dashed curve is the contribution only from the injected spicule material, while the green dashed curve is the contribution of the loop material excluding the injected spicule. The black curve is the sum of both the spicule and non-spicule intensities. {Time integrated intensities over the 190 s required for the shock to reach the right footpoint are indicated. Even though the emission from the injected spicule material decreases rapidly, it is so much brighter than the shock heated pre-existing material that it contributes more to the time integration.} }
	 \label{emission_evolution_fe12_fe14}
\end{figure*}

At $t=10$~s, the emission in both lines comes primarily from the injected hot plasma (orange region). On the other hand, at $t=70$~s it comes primarily from the shock heated plasma (red region). The transition happens very early on. Shortly after the injection of the hot material stops ($t = 10$ s), emission from the shock heated material starts dominating the total emission from the loop. This is evident in the time evolution plot of the loop-integrated emission in Figure~\ref{emission_evolution_fe12_fe14}. Shown are the intensities that would be observed by EIS, assuming that the loop has a cross section equal to the pixel area and that all of the loop plasma is contained within a single pixel. This corresponds to a loop that has been straightened along the line of sight and crudely represents a line of sight passing through an arcade of similar, out of phase loops. The black curve shows the evolution of the total emission contributed by the spicule and pre-existing plasma. Subtracting the spicule component (red curve) from the total gives the evolution of the emission coming solely from the pre-existing (non-spicule) loop material (green curve). Soon after the hot tip of the spicule completes its entry into the loop (at $t=10$~s), the emission from the spicule falls off rapidly. This is because the hot material at the spicule tip cools rapidly as it expands in the absence of any external heating. It is far too faint to make a significant contribution to the observed coronal emission, as emphasized earlier by \cite{Klimchuk_2012} and \cite{Klimchuk_2014}.

\begin{figure*}
\centering
	 \includegraphics[width=0.48 \textwidth,angle=0]{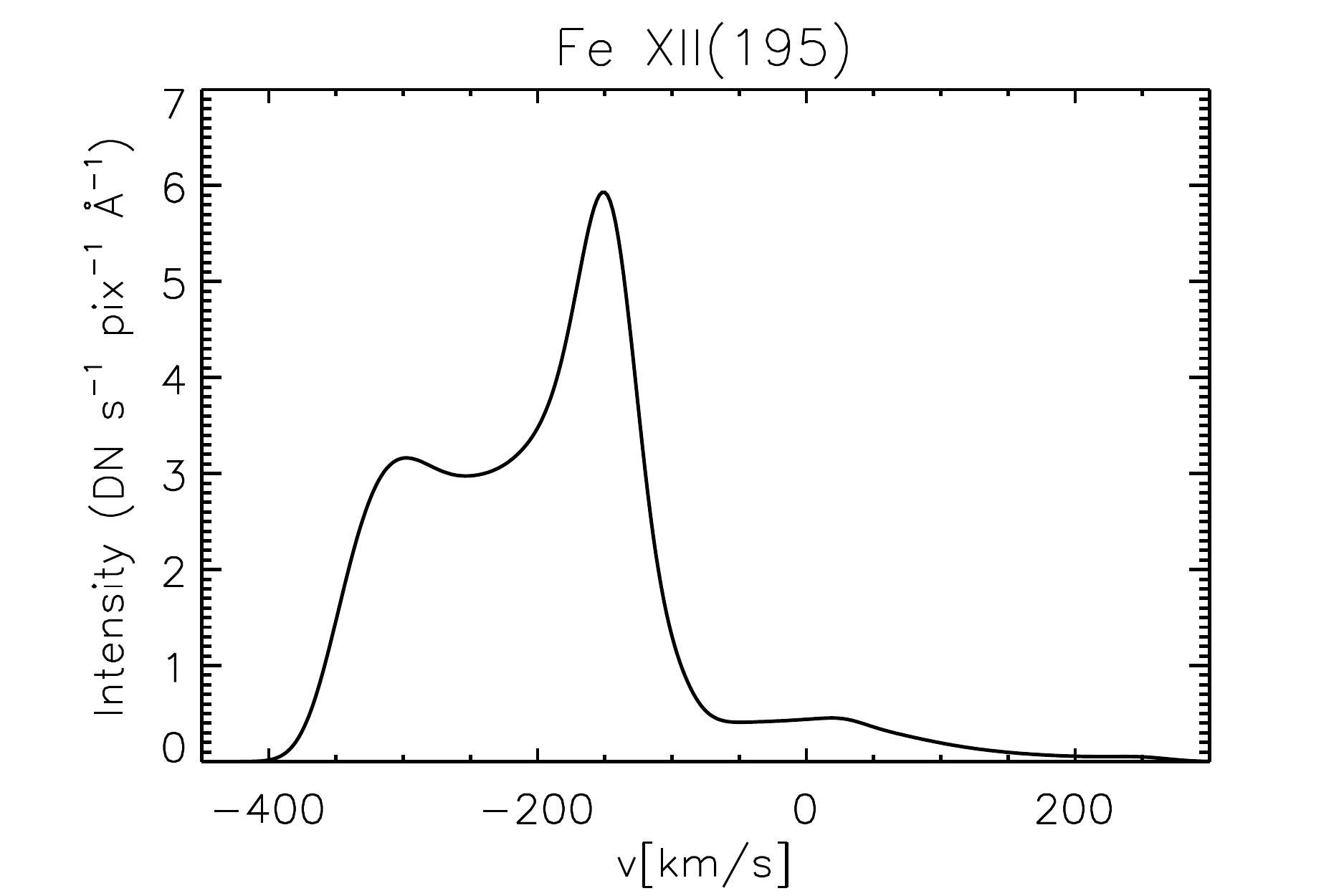}
	  \includegraphics[width=0.48 \textwidth,angle=0]{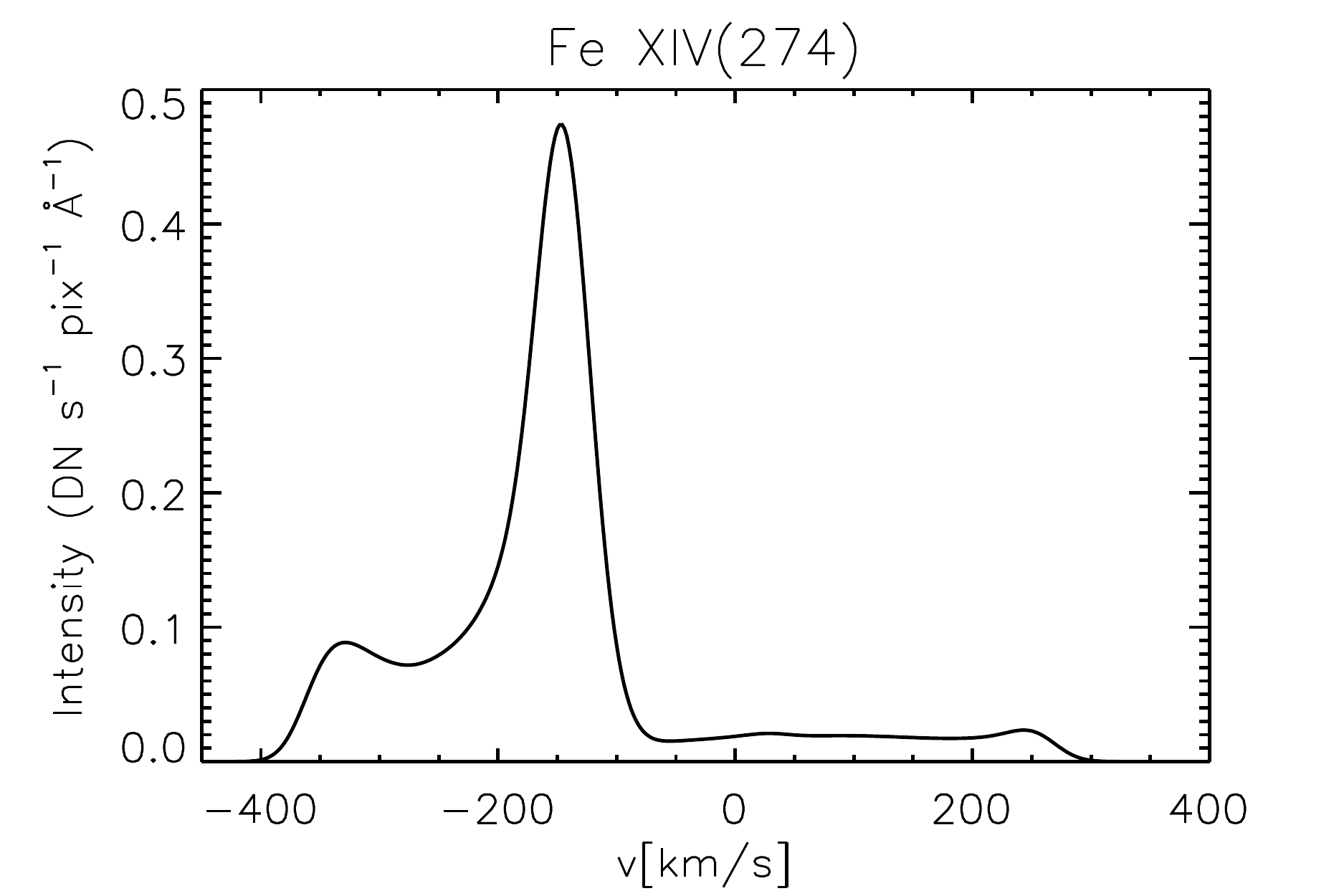}
	 \caption{Spatially and temporally averaged synthetic spectral line profiles of Fe XII and Fe XIV for Run1. The line profiles are averaged over the full length of the loop and over the 190 s needed for the shock to reach the other end of the loop. Spectral line profiles are highly blue-shifted, and the shapes do not agree with the observed asymmetric line profiles.}
	 \label{spectral_line_fe12_fe14}
\end{figure*}

For better comparison with observations, we construct synthetic spectral line profiles. The methodology is explained in Appendix~\ref{append:Forward_Modelling}. To construct these profiles, we imagine that the loop lies in a vertical plane and is observed from above. We account for the semi-circular shape when converting velocities to Doppler shifts. We then integrate the emission over the entire loop and distribute it uniformly along the projection of the loop onto the solar surface. We assume a cross section corresponding to an EIS pixel, and thereby obtain a spatially averaged EIS line profile for loop.  Finally, a temporal average is taken over the time required for the shock to travel to the other end of the loop ($\approx 190$~s in this case).  Such spatially and temporally averaged line profiles from a single loop (e.g., Figure~\ref{spectral_line_fe12_fe14}) is equivalent to an observation of many unresolved loops of similar nature but at different stages of their evolution \citep{Patsourakos_2006,Klimchuk_2014}. 

Asymmetric coronal line profiles with blue wing enhancement are manifestations of mass transport in the solar corona. Type II spicules are often suggested to be associated with such a mass transport mechanism ~\citep{Pontieu_2009, Pontieu_2011, Martinez_2017}. However, the extreme non-Gaussian shapes of the simulated Fe XII and Fe XIV  line profiles (Figure~\ref{spectral_line_fe12_fe14}) are significantly different from observed shapes~\citep{Pontieu_2009, Tian_2011, Tripathi_2013}. Also, the very large blue shifts are inconsistent with observations. Observed Doppler shifts of coronal lines tend to be slower than  $5$ km s$^{-1}$ in both active regions ~\citep{Doschek_2012,Tripathi_2012} and quiet Sun ~\citep{Chae_1998,Peter_1999}. In contrast, a shift of $150$ km s$^{-1}$ is evident in the simulated spectral lines (Figure~\ref{spectral_line_fe12_fe14}).

Our simulation is not reliable after the shock reaches the right boundary of the model. Because of rigid wall boundary conditions, it reflects in an unphysical manner. One might question whether the emission after this time could dramatically alter the predicted line profiles. We estimate the brightness of this neglected emission using the loop temperature profile shortly before the shock reaches the chromosphere at $t = 190$~s. The temperature peaks at the shock, and there is strong cooling from thermal conduction both to the left (up the loop leg) and, especially, to the right (down the loop leg). We estimate the cooling timescale according to:

\begin{equation}
    \tau_{cond}= \frac{21}{2}\frac{k_{B}n_{e}l^{2}}{\kappa_{0\parallel}T^{5/2}}~,
\end{equation} 

where, $k_{B}$ is the Boltzmann constant, $\kappa_{0\parallel}$ is the coefficient of thermal conductivity along the field lines, $T$ is the temperature at the shock, $n_{e}$ is the electron number density behind the shock, and $l$ is the temperature scale length. We do this separately using the scale lengths on both sides of the shock, obtaining $\tau_{cond}=1290$~s and {7}~s for the left and right sides, respectively. Radiative cooling is much weaker and can be safely ignored. We estimate the integrated emission after $t = 190$~s by multiplying the count rate at that time by the longer of the two timescales, thereby obtaining an upper limit on the neglected emission in our synthetic line profiles. The result is $10565$ DN pix$^{-1}$ for Fe XII and $2206$ DN pix$^{-1}$ for Fe XIV. These are about $0.97$ and $2.76$ times the temporally integrated emission before this time, for Fe XII and Fe XIV, respectively.
The factors are much smaller using the shorter cooling timescale. Even the large factors do not qualitatively alter our conclusions. The profile shapes and Doppler shifts would still be much different from observed. The conclusions we draw below are also not affected by neglecting the emission after the shock reaches the right footpoint.

\subsection{Comparison with observations}

We now estimate the spicule occurrence rate that would be required to explain the observed coronal intensities from active regions and quiet Sun. We have already seen that, in the absence of any external (coronal) heating, the hot material at the tip of the spicule cools down rapidly.  However, we are concerned here with the total emission, including that from pre-existing material that is heated as the spicule propagates along the loop. Consider a region of area $\mathcal{A}_{reg}$ on the solar surface, large enough to include many spicules. If the spatially averaged occurrence rate of spicules in this region is $\mathcal{R}$ (cm$^{-2}$ s$^{-1}$), then one may expect $\mathcal{N}_{reg} = \mathcal{R}\tau\mathcal{A}_{reg}$ spicules to be present at any moment, where $\tau$ is the typical spicule lifetime. Since we are averaging over large areas, the orientations of the spicule loops does not matter, and we can treat the loops as straightened along the line of sight, as done for Figure 5. If $\mathcal{I}_{sp}$ (DN s$^{-1}$ pix$^{-1}$) is the temporally averaged intensity of such a loop (the full loop intensity divided by 190 s in Fig. 5), then the expected intensity from a corona that only contains spicule loops is $\mathcal{I}_{obs} = \mathcal{N}_{reg}\mathcal{I}_{sp}\mathcal{A}_{sp}/\mathcal{A}_{reg} = \mathcal{I}_{sp}\mathcal{R}\tau\mathcal{A}_{sp}$,
where $\mathcal{A}_{sp}$ is the cross-sectional area of the loop.

\begin{figure*}
\centering
	 \includegraphics[width=0.48 \textwidth,angle=0]{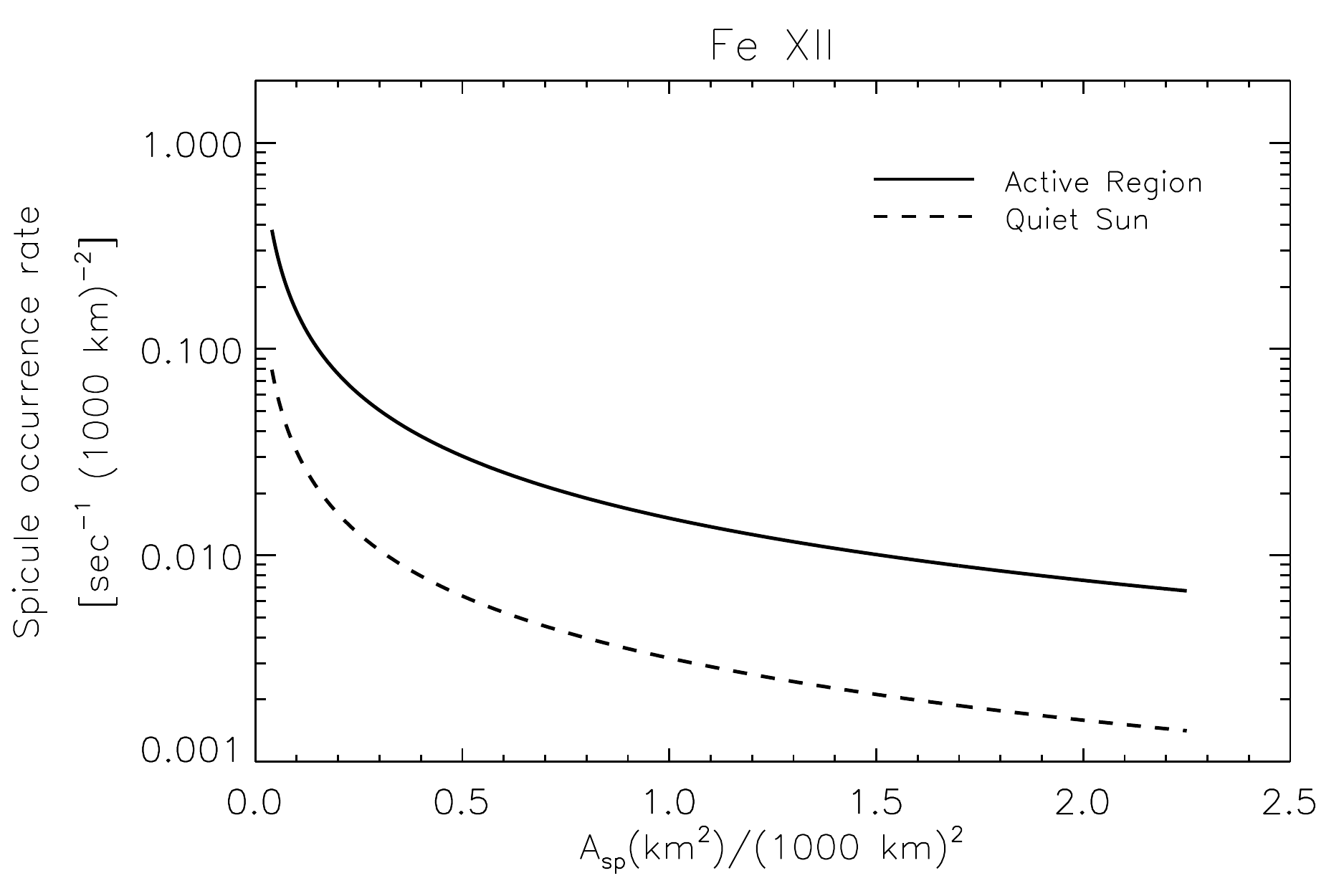}
	 \includegraphics[width=0.48 \textwidth,angle=0]{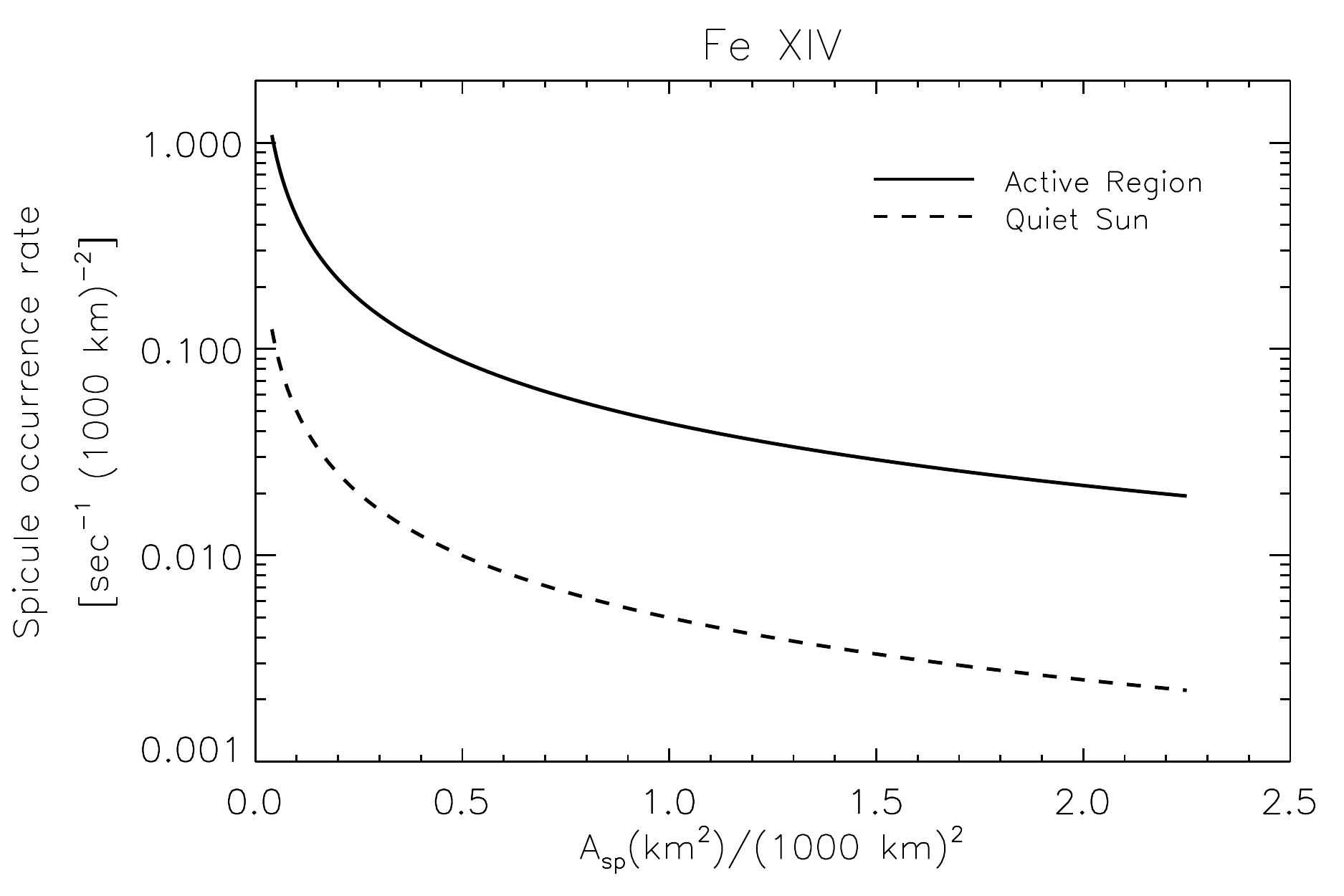}
	 \caption{The required occurrence rate of spicules to explain the typically observed Fe XII and Fe XIV intensities for active regions and quiet Sun (curves are derived from Run1). The occurrence rate is plotted against the cross-sectional area of spicules. }
	 \label{count_fe12_fe14}
\end{figure*}

The typical intensities ($\mathcal{I}_{obs}$) observed by EIS in active regions and quiet Sun are, respectively, $162$ and $34$ DN s$^{-1}$ pix$^{-1}$ in Fe XII (195 \AA) and $35$ and $4$ DN s$^{-1}$ pix$^{-1}$ in Fe XIV (274 \AA) ~\citep{Brown_2008}. On the other hand, the temporally averaged intensities from our simulation ($\mathcal{I}_{sp}$) are $56.36$ and $4.22$ DN s$^{-1}$ pix$^{-1}$ for Fe XII and Fe XIV, respectively. Considering $\tau$ to be $190$~s, the time it takes for the shock to travel across the loop, we derive an occurrence rate ($\mathcal{R}$) of spicules as a function of their cross-sectional area ($\mathcal{A}_{sp}$). Results are shown in Figure~\ref{count_fe12_fe14} for the two lines. 

\begin{figure*}
\centering
	  \includegraphics[width=0.48 \textwidth,angle=0]{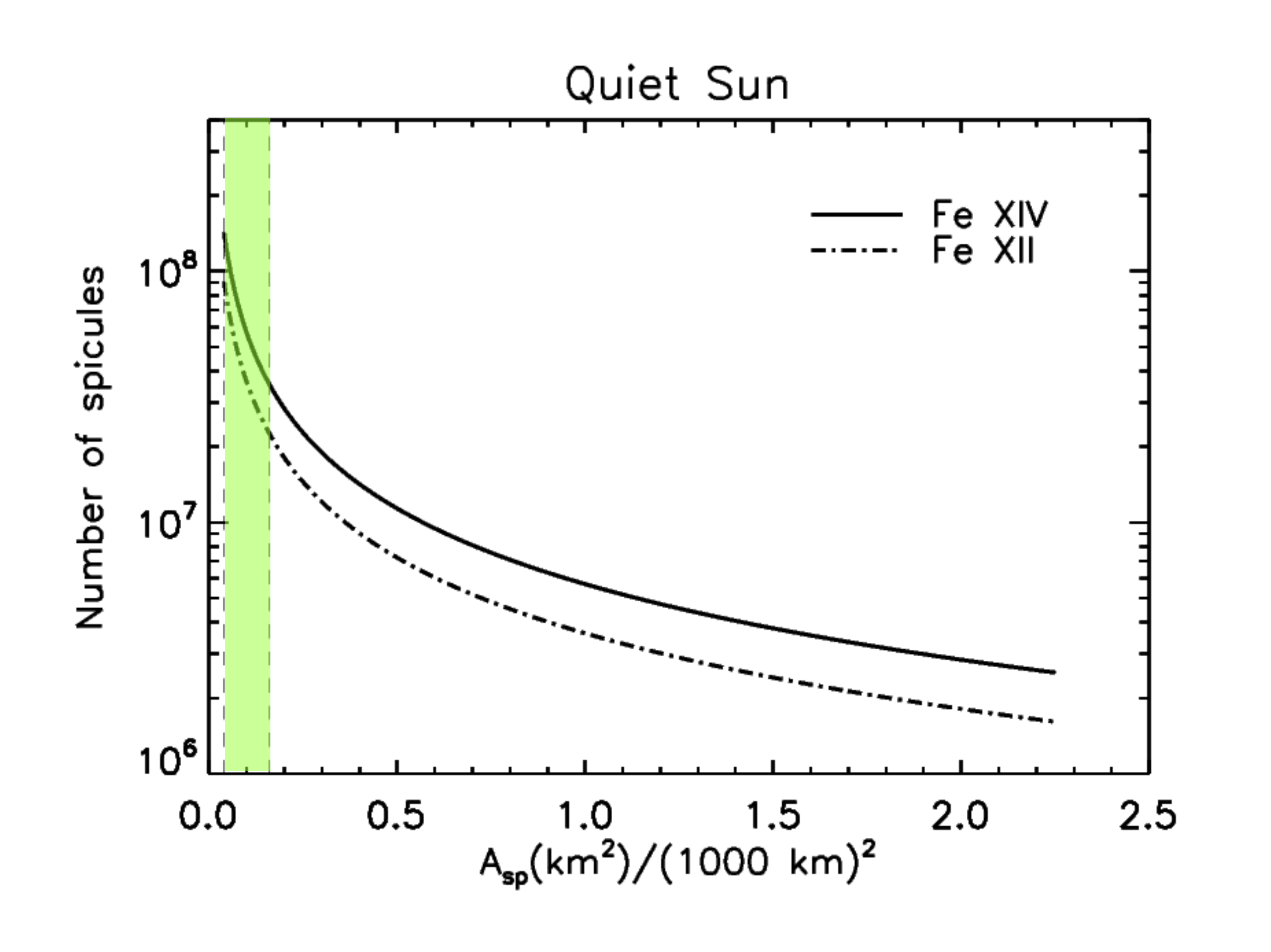}
	   \includegraphics[width=0.48 \textwidth,angle=0]{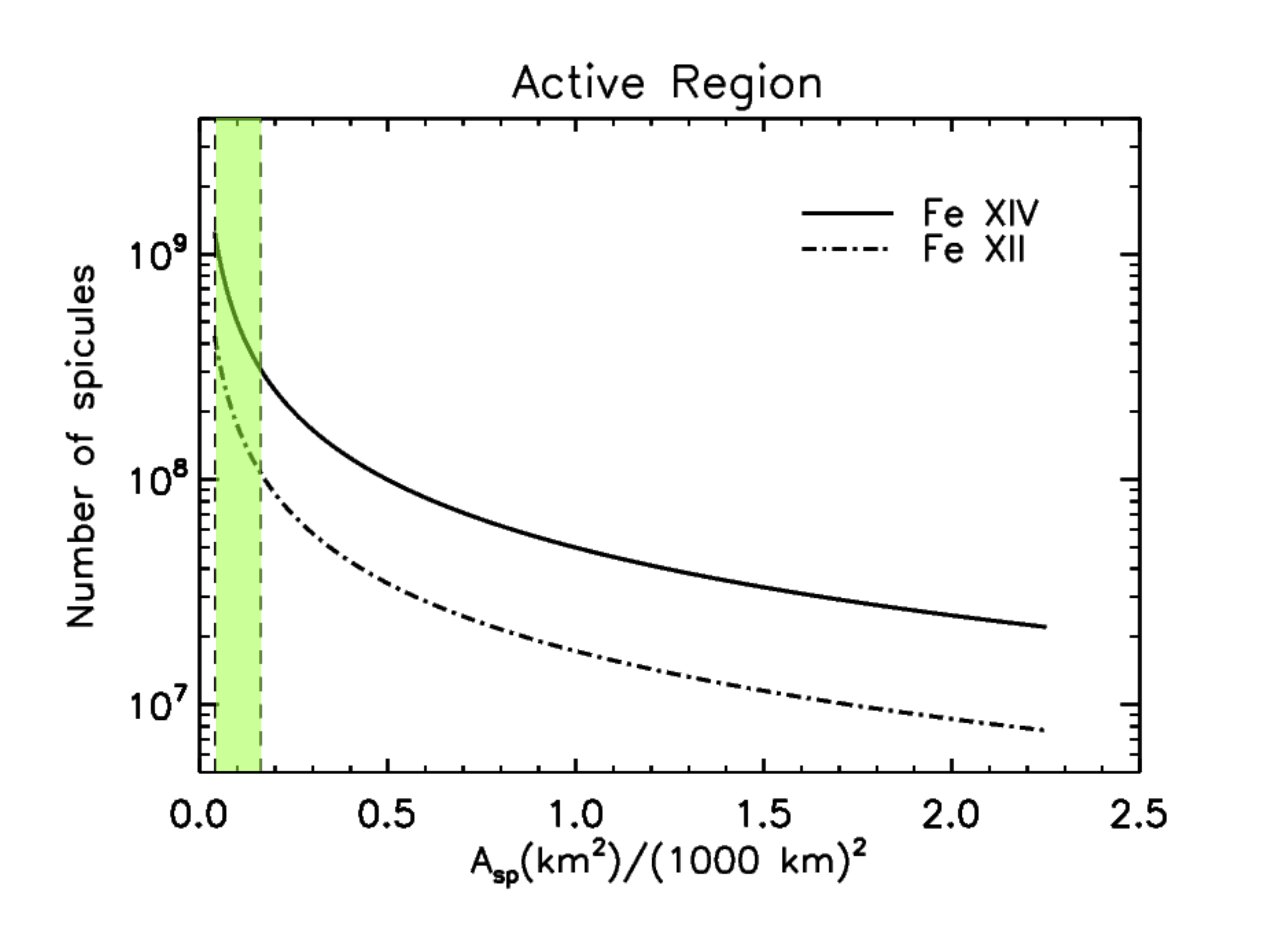}
	 \caption{Required number of spicules at any given time on the Sun as a function of spicule cross-sectional area, estimated from the spicule occurrence rate (derived using Run1). Spicules having widths of $200 - 400$ km are fairly common. The required number of spicules having widths in this range is marked by the shaded region in both the plots. }
	 \label{count_QS_AR}
\end{figure*}

Following our earlier logic, we may also argue that at any given time there are $\mathcal{N}_{\odot}=\mathcal{R}\tau \mathcal{A}_{\odot}$ spicules on the solar disk, where $\mathcal{A}_{\odot}$ is the area of the solar disk. Using the estimated value of the occurrence rate of the spicules ($\mathcal{R}$), and taking $\tau$ to be $190$~s as before, the number of spicules on the solar disk is related to the other quantities as per $\mathcal{N}_{\odot} = (\mathcal{I}_{obs}/\mathcal{I}_{sp})(\mathcal{A}_{\odot}/\mathcal{A}_{sp})$. This formula represents $\mathcal{N}_{\odot}$ as a function of the spicule cross-sectional area $\mathcal{A}_{sp}$ (Figure~\ref{count_QS_AR}). Considering the fact that the typical observed widths of spicules lie between $200-400$~km~\citep{Pereira_2011}, we find that the full disk equivalent number of spicules required to explain the observed intensities exceeds $10^7$ in the quiet Sun and $10^8$ in active regions, as indicated by the green shaded region in Figure~\ref{count_QS_AR}. However, observational estimations for the number of spicules on the disk vary between $10^5$~\citep{Sterling_2016} and $2 \times 10^7$~\citep{Judge_2010}. There is a large discrepancy. Far more spicules than observed would be required to produce all the observed coronal emission. For the quiet Sun, $100$ times more spicules would be needed, while for active regions,  $10-10^3$ times more would be needed. These are lower limits based on Run1. Our other simulations imply even greater numbers of spicules (see Table~\ref{tab:table2}).

We should mention here that the larger the height the spicule rises, the longer the time it compresses the ambient material, and thus the brighter the time averaged emission.  The spicules in our simulations with $150$ km s$^{-1}$ injection speed reach a height of about $23$~Mm, which is much larger than the typically observed spicule height ($\sim 10$~Mm). Therefore, we are likely to overestimate the emission coming from spicule loops, and so the discrepancy between the required and observed number of spicules is even greater. It should also be noted that the values estimated by ~\cite{Sterling_2016, Judge_2010} consider both type I \& II spicules. The discrepancy thus increases further if one considers type II spicules alone. 

Analysis of our simulated observations thus suggests that spicules contribute a relatively minor amount to the emission and thermal energy of the corona. Through the generation of shocks, they may heat the local plasma, but that too cools down rapidly due to expansion and thermal conduction. Therefore, synthetic spectra derived from our simulation show a high discrepancy with observed spectra. However, this does not rule out the possibility of spicules contributing significantly to the coronal mass. The ejected spicule material may still get heated in the corona through some other heating mechanism -- a source that exceeds the initial thermal and kinetic energy of the spicule. However, observational evidence of such a process is still lacking. Analyzing the excess blue wing emission of multiple spectral lines hotter than $0.6$~MK, \cite{Tripathi_2013} have concluded that the upward mass flux is too small to explain the mass of the active region corona. Their observations indicate that spicules hotter than $0.6$~MK are not capable of providing sufficient mass to the corona. 

So far, we have allowed our spicules to propagate within a warm ($T =0.5$ MK), relatively low density loop in order to determine whether they, by themselves, can explain the observed hot emission. Our simulations indicate that this is not viable. Therefore, there must be some other heating mechanisms at play that produce the hot, dense plasma. Setting aside the issue of heating mechanisms, in the following section we simply test the response of a spicule in a hot and dense flux tube.

\subsection{Spicule propagation in a hot loop}

We have considered a static equilibrium loop with apex and footpoint temperatures of approximately $2$  and $0.02$ MK, respectively. A spicule with a tip temperature of $2$ MK followed by a cold, dense material with temperature $0.02$ MK is injected with a velocity of $150$ km s$^{-1}$ from the bottom boundary, similar to our previous spicules. The velocity profile of the injected spicule is the same as shown in Figure~\ref{pulse}. The injected spicule generates a shock that takes about $180$~s to traverse the loop. 

The spatio-temporal averaged spectral line profiles are obtained following the method described in Appendix~\ref{append:Forward_Modelling}. However, in this case, because of the high background temperature, the loop itself emits significantly in the Fe XII and Fe XIV coronal lines. We consider the situation where the line of sight passes through many loops. Some contain spicules and some are maintained in the hot equilibrium state. We adjust the relative proportions to determine what combination is able to reproduce the observed red-blue (RB) profile asymmetries, which are generally $< 0.05$ \citep{Hara_2008, Pontieu_2009, Tian_2011}. For an asymmetry of $\approx 0.04$, we find that the ratios of spicule to non-spicule strands are $1:150$ for the Fe XII line and $1:72$ for the Fe XIV line. Again the conclusion is that spicules are a relatively minor contributor to the corona overall, though they are important for the loops in which they occur.

\begin{figure}
\centering
	 \includegraphics[width=0.48 \textwidth,angle=0]{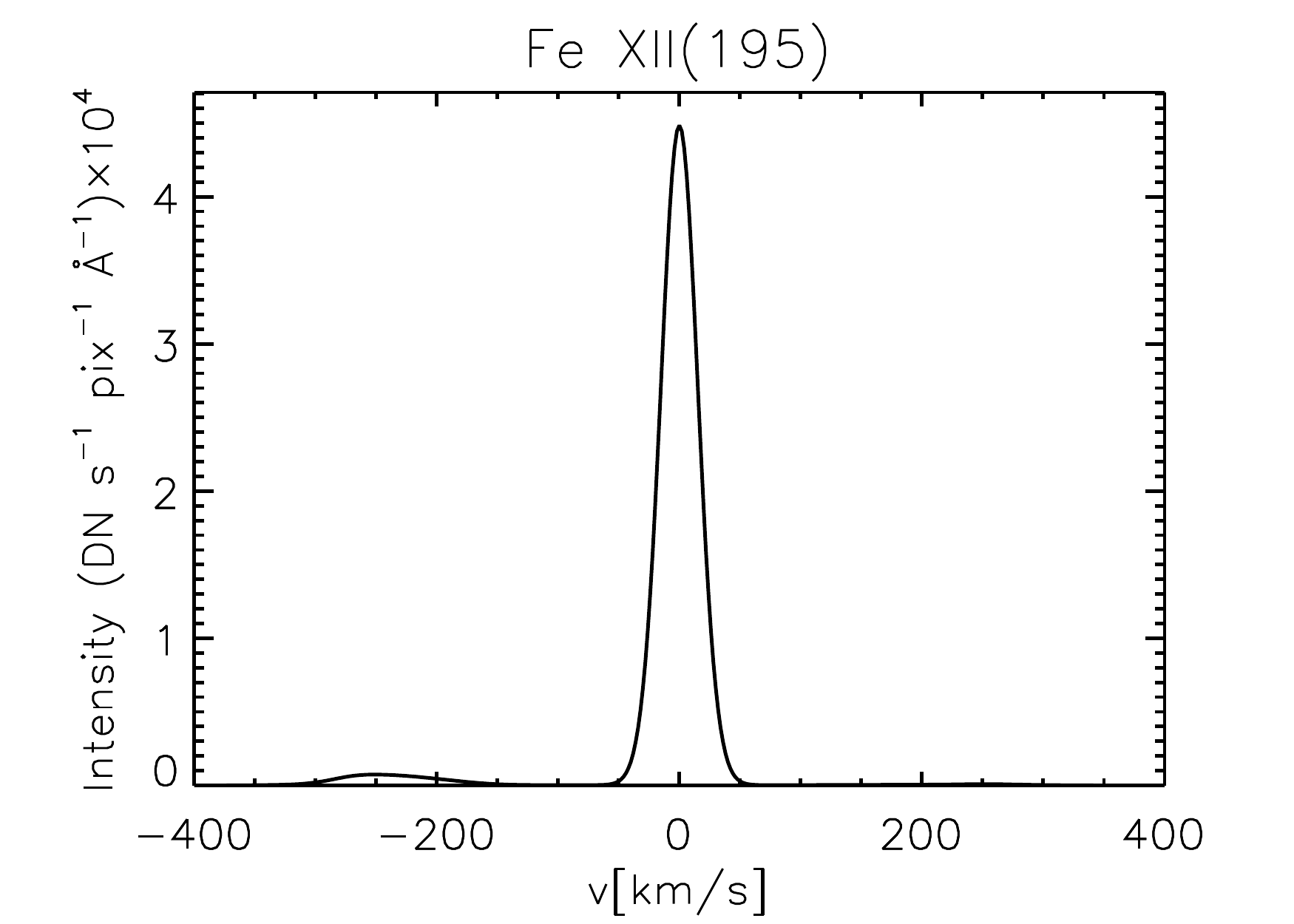}
	  \includegraphics[width=0.48 \textwidth,angle=0]{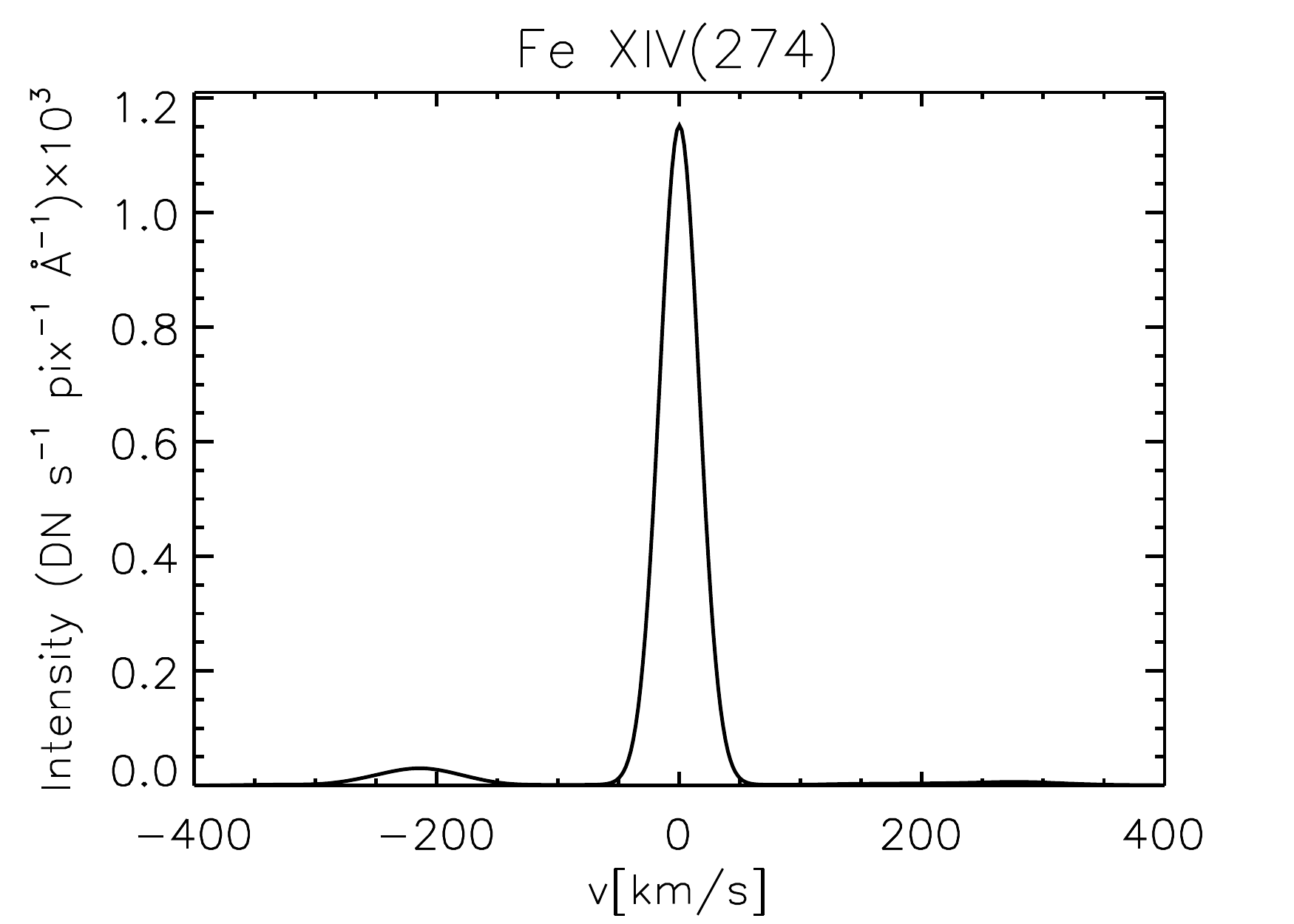}
	 \caption{Spatially and temporally averaged synthetic spectral line profiles of Fe XII and Fe XIV for the case when a spicule (with tip temperature of $2$~MK) is injected in a hot loop with apex temperature $2$~MK. The line profiles are averaged over $180$~s (the time taken by the shock to traverse the loop).}
	 \label{spectral_line_fe12_fe14_with_bck}
\end{figure}

\begin{deluxetable*}{ccccccccc}
\tablenum{2}
\tablecaption{Summarizing the number of spicules (width $\sim 300$ km) required to explain the quiet Sun and active region intensities as predicted from the test runs.}\label{tab:table2}
\tablecolumns{9}
\tablewidth{0pt}
\tablehead{
\colhead{Run} & \colhead{T$_{tip}$} & \colhead{v} & \multicolumn2c{Loop integrated counts} & \multicolumn2c{Quiet Sun} & \multicolumn2c{Active Region} \\
\colhead{} &\colhead{(MK)} & \colhead{(km s$^{-1}$)} & \multicolumn2c{(DN s$^{-1}$ pix$^{-1}$)} & 
\multicolumn2c{(Required number of spicules)}  & \multicolumn2c{(Required number of spicules)} \\
\colhead{} & \colhead{} & \colhead{} & \colhead{Fe XII} &  \colhead{Fe XIV}  & \colhead{Fe XII} & \colhead{Fe XIV} & \colhead{Fe XII} & \colhead{Fe XIV} 
}
\startdata
1&2    & 150  & 0.6405 & $4.8\times10^{-2}$  & $4.02\times10^{7}$ & $6.31\times10^{7}$ & $1.92\times10^{8}$ & $5.52\times10^{8}$\\
2&2    & 50   & 0.2336 & $1.47\times10^{-2}$ & $1.1\times10^{8}$ & $2.06\times10^{8}$ & $5.25\times10^{8}$ & $1.8\times10^{9}$ \\
3&1    & 150  & $6.7\times10^{-2}$  & $1.3\times10^{-3}$  & $3.86\times10^{8}$ & $2.26\times10^{9}$ & $1.84\times10^{9}$ & $1.98\times10^{10}$\\
4&1    & 50   & $5.9\times10^{-3}$ & $7.81\times10^{-6}$ & $4.3\times10^{9}$ & $3.8\times10^{11}$ & $2.06\times10^{10}$ & $3.4\times10^{12}$ \\
5&0.02 & 150  & $4.4\times10^{-4}$  & $1.5\times10^{-7}$ & $5.89\times10^{10}$ & $2.02\times10^{13}$ & $2.8\times10^{11}$ & $1.77\times10^{14}$ \\
6&0.02 & 50   & $1.02\times10^{-7}$ & $1.1\times10^{-13}$ & $2.5\times10^{14}$ & $2.7\times10^{19}$ & $1.2\times10^{15}$ & $2.4\times10^{20}$ \\
\enddata
\end{deluxetable*}
\section{Summary and discussion}\label{sec:summary}
The solar atmosphere displays a wide variety of spicules with different temperatures and velocities. It has been suggested that type II spicules are a major source of coronal mass and energy~\citep{Pontieu_2007,Pontieu_2009,Pontieu_2011}. In this work, we numerically investigate the role of spicules in producing observed coronal emissions. In particular, we examine whether, in the absence of any external heating, the hot tips of the spicules and the shock-heated ambient plasma can explain the observed coronal emission. For this, we inject spicules with different temperatures and velocities into a coronal loop in static equilibrium. We choose a relatively cool equilibrium so that the loop does not itself produce appreciable emission in the absence of a spicule. Each of our injected spicules consists of a hot tip followed by a cold body. We consider three different temperatures for the hot tips, viz., $2$, $1$ and $0.02$~MK, while the cold, dense chromospheric plasma that follows the tip has a temperature of $0.02$~MK. Six different simulations are run by injecting each of these spicules with an initial velocity of either $50$ km s$^{-1}$ or $150$ km s$^{-1}$ (see Table~\ref{tab:table1}). We also have constructed spectral line profiles and estimated the spicule occurrence rate required to explain the observed intensities from the quiet Sun and active regions. Our main results are summarized as follows.

\paragraph{Shock formation during spicule propagation} All six runs described above suggest the formation of shocks due to the injection of spicule material into the  coronal flux tubes. The shocks are stronger when the temperature differences and therefore pressure differences with the ambient plasma are higher. Table~\ref{tab:table1} shows the variation of the compression ratio (measure of shock strength) with changing temperature of the spicule tip. The nature of the shock depends on the tip temperature. Spicules with a hotter tip produce a pressure-driven shock that propagates with a speed larger than the injection speed. Spicules with a cold tip (i.e., $T_{tip} = 0.02$ MK) produce a piston-driven shock which propagates with a speed close to the injection speed. The intensities and shapes of spectral line profiles depend on the nature of the shock. The formation of shocks during spicule injection agrees well with previous studies~\citep{Petralia_2014, Martinez_2018}.

\paragraph{Rapid cooling of the hot spicule tip} Our simulations show that, in the absence of any external heating, the hot tip of the spicule cools rapidly before reaching a substantial coronal height. Consequently, the tip emission from coronal lines like Fe XII (195 \AA) and Fe XIV (274 \AA) is short lived (Figure~\ref{emission_evolution_fe12_fe14}) and confined to low altitudes. The result is consistent with earlier studies by~\cite{Klimchuk_2012} and \cite{Klimchuk_2014}.

\paragraph{Relative emission contributions of hot tip and shock heated plasma} Our simulations show that the pre-existing material in the loop gets heated through shock compression and thermal conduction. However, the time-integrated emission from this heated pre-existing material is less than that from the hot tip, as shown in Figure~\ref{emission_evolution_fe12_fe14}. The tip plasma is hot for a much shorter time, but it is inherently much brighter because of the greater densities (it is injected in a dense state).

\paragraph{Line profile discrepancies} The shapes of our synthetic spectral line profiles show significant discrepancies with observations. The simulated profiles are highly non-Gaussian and far more asymmetric than observed. A strong blue shift ($\sim 150$ km s$^{-1}$) of the synthetic lines is also inconsistent with the mild Doppler shifts ($< 5$ km s$^{-1}$) observed in the quiet Sun and active regions.

\paragraph{Excessive number of spicules required to explain observed intensities} The spatially and temporally averaged intensities from our simulations (Figures~\ref{spectral_line_fe12_fe14}) imply that far more spicules are required to reproduce the observed emission from the solar disk than are observed (Figure~\ref{count_QS_AR}). The discrepancies are up to a factor of $100$ for the quiet Sun and factors of $10-10^3$ for active regions. These factors apply specifically to Run1, where a spicule with a $2$ MK tip is ejected at a velocity of $150$ km s$^{-1}$. As listed in Table \ref{tab:table2}, the loops in our other simulations with different combinations of tip temperature and ejection velocity are fainter, and therefore more of them would be required to reproduce the observed disk emission, exacerbating the discrepancy.

\paragraph{Ratio of loops with and without spicules} Under the assumption that the corona is comprised of hot loops with and without spicule ejections, red-blue spectral line asymmetries similar to those observed ($0.04$) require far more loops without spicules than with them. The spicule to non-spicule loop number ratio is $1:150$ for the FeXII line and $1:72$ for the Fe XIV line. 

Our simulations indicate that spicules contribute a relatively minor amount to the mass and energy of the corona. Such a claim had already been made by \cite{Klimchuk_2012}, where it was shown that hot tip material rapidly expanding into the corona is unable to explain the observed coronal emission. However, a bodily ejection of the spicule was not considered, and the emission from ambient material effected by the expansion was not rigorously investigated (though see Appendix B in that paper). Later, \cite{Petralia_2014} argued that the shock-heated material in front of an ejected cold spicule might be erroneously interpreted as ejected hot material. They did not compare the brightness of the shock-heated material with coronal observations. Our numerical simulations improve on both of these studies. We show that neither the expanding hot tip nor the shock-heated ambient material of a bodily ejected spicule can reproduce coronal observations. A number of discrepancies exist. The existence of some coronal heating mechanism - operating in the corona itself - is required to explain the hot corona. It is not sufficient to eject hot (or cold) material into the corona from below. 

We emphasize that our conclusion does not rule out the possibility that waves may be launched into the corona as part of the spicule ejection process, or that new coronal currents may be created outside the flux tube in which the ejected material resides, as suggested by ~\cite{Martinez_2018}. Such waves and currents would lead to coronal heating and could explain at least some non-spicule loops. It seems doubtful, however, that this could explain the many non-spicule loops implied by observed line profile asymmetries. It seems that some type of heating unrelated to spicules must play the primary role in explaining hot coronal plasma.

\begin{acknowledgments}
We thank the anonymous referee for her/his comments to improve the clarity of the paper. SSM \& AS thank Dr. Jishnu Bhattacharyya for many useful discussions. Computations were carried out on the Physical Research Laboratory's VIKRAM cluster. JAK was supported by the Internal Scientist Funding Model (competed work package program) at Goddard Space Flight Center. \\
\end{acknowledgments}

\appendix
\section{Static equilibrium configuration from double relaxation method}\label{append:steady_state}
We inject spicules in a magnetic structure that is in static equilibrium. Such an equilibrium is achieved recursively, and the final equilibrium profile is obtained through two stages of relaxation. First, we obtain the density and temperature profiles by solving the hydrostatic and energy balance equations~\citep{Aschwanden_2002} assuming a steady and uniform background heating $Q_{bg}$. The CHIANTI radiative loss function $\Lambda(T)$ is used to describe the loop's radiation in the energy balance equation. The desired looptop temperature is achieved by adjusting the value of $Q_{bg}$. However, due to lack of exact energy balance, the temperature and density profiles derived in this way do not achieve a perfect equilibrium state. Rather these derived profiles are then used to calculate the final equilibrium loop profile , such that the resulting temperature profile never drops below the chromospheric temperature $T_{ch}$ ($2 \times 10^4$~K), and the system does not generate any spurious velocity either. In the following, we explain these two stages in detail.

\subsection{Heating and cooling in Relaxation-I:} 
Starting with the initial profiles described above, the loop is allowed to relax under gravity with the constant background heating $Q_{bg}$. To avoid numerical artifacts, from this stage onward, we smoothly reduce the radiative cooling of the chromosphere to zero over a narrow temperature range between $T_{ch}$ and $T_{min}$, where $T_{min} = 1.95 \times 10^4$~K is a conveniently chosen temperature slightly less than $T_{ch}$. This is achieved by the radiative loss function $\lambda(T)$, defined as
\begin{equation}\label{rad_relax1}
 \lambda(T) =
  \begin{cases}
    \Lambda(T), & \text{if}\, T \ge T_{ch} \\
    \left(\frac{T - T_{min}}{T_{ch} - T_{min}}\right) \Lambda(T_{ch}), & \text{if}\, T_{min} < T < T_{ch} \\
    0, & \text{if}\, T \le T_{min} \\
  \end{cases}~.
\end{equation}
Here $\Lambda(T)$ denotes the optically thin radiative loss function from CHIANTI. The modified function $\lambda(T)$ is plotted in Figure~\ref{chrom_heat_cool}. As the loop relaxes, material drains from the corona and accumulates at the footpoints. The resulting high density of the loop footpoints gives rise to excessive cooling and brings down the footpoint temperatures below $T_{min}$, along with generating short lived velocities. However, the loop eventually achieves a steady-state, and we use the enhanced footpoint density at that time ($n_{base}$) to estimate the additional heating required to keep the chromospheric temperature above $T_{min}$. This is carried out in the next relaxation stage.

\subsection{Heating and cooling in Relaxation-II:} 
\begin{figure}
\centering
	 \includegraphics[width=0.55 \textwidth,angle=0]{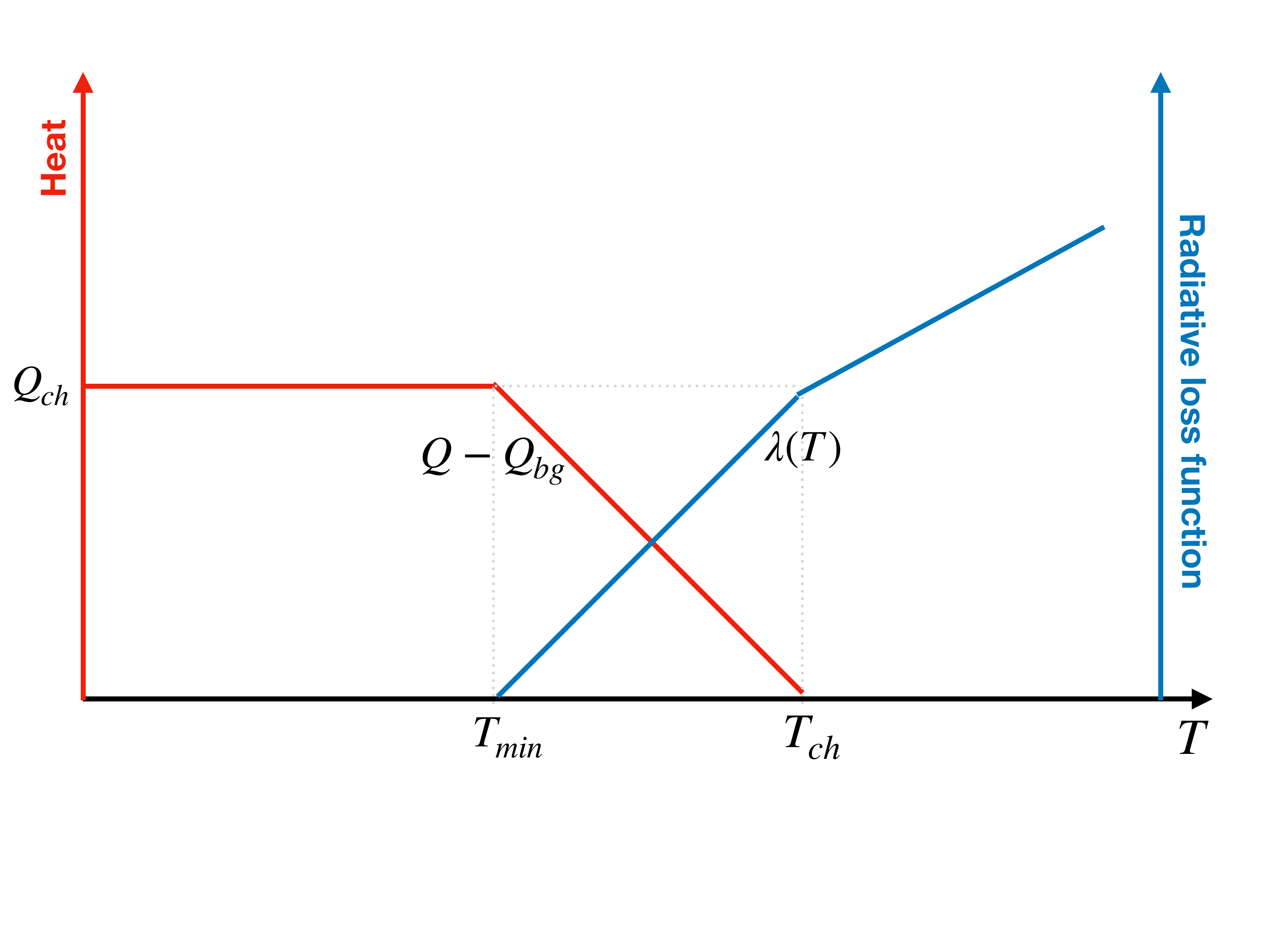}
	 \caption{Radiative loss function and excess heating implemented to prevent cooling of plasma below $T_{min}$. When the temperature of the loop is above $T_{ch}$, the radiative loss is given by the CHIANTI radiative loss function $\Lambda(T)$ (Equation~\ref{rad_relax1}). The only heating applied to the loop at these temperatures is the uniform background heating $Q_{bg}$. The radiative cooling smoothly goes to zero as the temperature approaches $T_{min}$ from above. In the temperature range $T_{min}$ to $T_{ch}$, depending on the plasma density and temperature, an additional heating is provided to the loop to balance the lost energy through radiative cooling. Below the temperature $T_{min}$, an additional heating $Q_{ch}$, proportional to $n_{ch}^2$, is applied to bring the plasma back $T_{ch}$.}
	 \label{chrom_heat_cool}
\end{figure}

Once again, we start with the initial density and temperature profiles from the beginning of the first stage. However, this time we apply additional heating in the chromosphere above the constant background heating $Q_{bg}$. This prevents the plasma from cooling below $T_{min}$ and instead lets it hover between $T_{ch}$ and $T_{min}$. The total heating function $Q$ is given by
\begin{equation}\label{heat_relax2}
 Q =
  \begin{cases}
    Q_{bg}, & \text{if}\, T \ge T_{ch} \\
    \left(\frac{n}{n_{base}}\right)^{2} Q_{ch} \left(\frac{T_{ch}-T}{T_{ch}-T_{min}}\right) + Q_{bg}, & \text{if}\, T_{min} < T < T_{ch} \\
    \left(\frac{n}{n_{base}}\right)^{2} Q_{ch} + Q_{bg}, & \text{if}\, T \le T_{min}
  \end{cases}~,
\end{equation}
where $Q_{ch} = n_{ch}^{2} \Lambda (T_{ch})$ is the heat required to balance the radiative losses from the footpoint plasma of the initial loop profile at temperature $T_{ch}$ and density $n_{ch}$. Figure~\ref{chrom_heat_cool} graphically depicts the radiative loss and heating functions that are maintained throughout the simulation.

\section{Variation of shock speed with height}\label{append:shock_speed}

\begin{figure}
\centering
	 \includegraphics[width=0.48 \textwidth,angle=0]{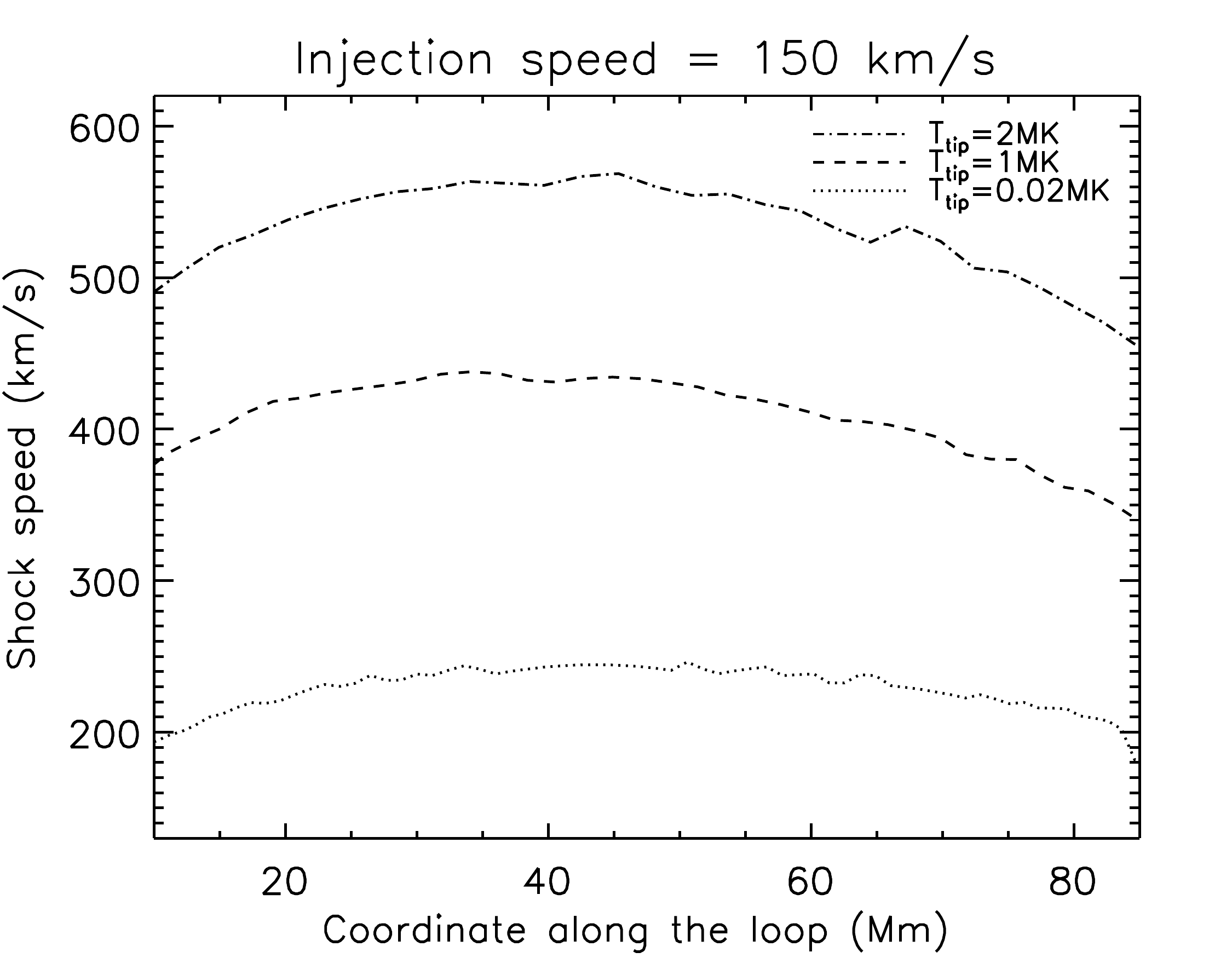}
	 \caption{Variation of the shock speeds as a function of loop coordinates for injection speed = $150$ km s$^{-1}$ and tip temperatures = $2$, $1$ and $0.02$ MK. }
	 \label{shock_speed}
\end{figure}

For a pressure driven shock, the shock's speed primarily depends on the pressure difference between the spicule's tip and the ambient medium in which it is propagating. Lower pressure close to the loop apex provides lesser resistance to the shock propagation, and hence the shock speed increases. On the other hand, high pressure close to the footpoints provides greater resistance and thus the shock speed reduces. For a better understanding, we track the shock front along the loop and derive its speed during its propagation. The shock front at any instant can be identified from the density jump moving ahead of the injected spicule material. To track it, we identify the jump in density at each time, which is also associated with the maximum temperature of the loop. Once the locations of the shock front along the loop are identified, a derivative of the same gives the instantaneous shock speed as a function of loop coordinates. Figure~\ref{shock_speed} shows the variation of shock speed as a function of loop coordinates for three different shocks, all ejected with velocity $150$ km s$^{-1}$ but with three different tip temperatures, viz. $2$, $1$ and $0.02$ MK. Though the shock speeds increase at the loop apex for all three shocks, velocity amplitudes depend on the injection temperatures and thus pressures. The larger the tip temperature, the higher the spicule tip pressure and hence larger is the shock speed. 

\section{Forward Modelling}\label{append:Forward_Modelling}

Spectral profiles provide a wealth of information about the plasma dynamics along the line of sight (LOS). Adapting the method outlined in~\cite{Patsourakos_2006}, synthetic spectral line profiles are constructed at each numerical grid cell using the cell's density, velocity and temperature. 
At any given time, $t$, and location along the loop, $s$, the line profile is
        \begin{equation}
         I(s,t) = \frac{I_{0}}{\sqrt{\pi} v_{\text{width}}}\exp\left[\frac{-(v - v_\text{shift})^{2}}{v_\text{width}^{2}}\right]~,
        \end{equation}
where $I_{0}$ is the amplitude, $v_\text{shift}$ is the Doppler shift, and $v_\text{width}$ is the thermal line width. The amplitude is given by 
        \begin{equation}
         I_{0}(s,t) = n_{e}^{2} G(T)ds~,
        \end{equation}
where $n_{e}$, $T$ and $ds$ denote the electron number density, temperature, and length of the cell. The contribution function $G(T)$ for the line is taken from the CHIANTI atomic data base \citep{chianti}. The Doppler shift equals the line of sight velocity of the cell,
        \begin{equation}
        v_\text{shift} = v_\text{los}~
        \end{equation}
in wavelength units, and the thermal width is given by 
        \begin{equation}
         v_\text{width} = \sqrt{\frac{2k_{B}T}{m_{ion}}}~,
        \end{equation}
where $m_\text{ion}$ is the mass of the ion. 
        
Once the line profile at each grid point is constructed, spatial averaging is performed by summing the profiles along the loop and dividing by its projected length assuming that it lies in a vertical plane and is viewed from above: 
\begin{equation}
    \langle I(t) \rangle_{\text{spatial}} = \frac{\pi}{2L}\sum_{s} I(s,t) \times d
\end{equation}
where $L$ is the loop length and $d$ is the pixel dimension. The loop is assumed to have a cross section of $d^2$. Finally the spatially averaged line profiles are temporally averaged over a time $\tau$, which is taken to be the travel time of the shock along the loop; this yields
\begin{equation}
    \langle I \rangle_{\text{spatial, temporal}} = \frac{1}{\tau}\sum_{t}\langle I(t) \rangle_{\text{spatial}}
\end{equation}


\bibliography{spicule}

\begin{thebibliography}{42}
\expandafter\ifx\csname natexlab\endcsname\relax\def\natexlab#1{#1}\fi

\bibitem[{{Aschwanden} \& {Schrijver}(2002)}]{Aschwanden_2002}
{Aschwanden}, M.~J., \& {Schrijver}, C.~J. 2002, \apjs, 142, 269

\bibitem[{{Athay} \& {Holzer}(1982)}]{Athay_1982}
{Athay}, R.~G., \& {Holzer}, T.~E. 1982, \apj, 255, 743

\bibitem[{{Bradshaw} \& {Klimchuk}(2015)}]{Bradshaw_2015}
{Bradshaw}, S.~J., \& {Klimchuk}, J.~A. 2015, \apj, 811, 129

\bibitem[{{Brown} {et~al.}(2008){Brown}, {Feldman}, {Seely}, {Korendyke}, \&
  {Hara}}]{Brown_2008}
{Brown}, C.~M., {Feldman}, U., {Seely}, J.~F., {Korendyke}, C.~M., \& {Hara},
  H. 2008, \apjs, 176, 511

\bibitem[{{Cargill} {et~al.}(2022){Cargill}, {Bradshaw}, {Klimchuk}, \&
  {Barnes}}]{Cargill_22}
{Cargill}, P.~J., {Bradshaw}, S.~J., {Klimchuk}, J.~A., \& {Barnes}, W.~T.
  2022, \mnras, 509, 4420

\bibitem[{{Chae} {et~al.}(1998){Chae}, {Yun}, \& {Poland}}]{Chae_1998}
{Chae}, J., {Yun}, H.~S., \& {Poland}, A.~I. 1998, \apjs, 114, 151

\bibitem[{Culhane {et~al.}(2007)Culhane, Harra, James, Al-Janabi, Bradley,
  Chaudry, Rees, Tandy, Thomas, Whillock, Winter, Doschek, Korendyke, Brown,
  Myers, Mariska, Seely, Lang, Kent, Shaughnessy, Young, Simnett, Castelli,
  Mahmoud, Mapson-Menard, Probyn, Thomas, Davila, Dere, Windt, Shea, Hagood,
  Moye, Hara, Watanabe, Matsuzaki, Kosugi, Hansteen, \& Wikstol}]{Culhane_2007}
Culhane, J.~L., Harra, L.~K., James, A.~M., {et~al.} 2007, Solar Physics, 243,
  19

\bibitem[{{De Pontieu} {et~al.}(2009){De Pontieu}, {McIntosh}, {Hansteen}, \&
  {Schrijver}}]{Pontieu_2009}
{De Pontieu}, B., {McIntosh}, S.~W., {Hansteen}, V.~H., \& {Schrijver}, C.~J.
  2009, \apjl, 701, L1

\bibitem[{{De Pontieu} {et~al.}(2007){De Pontieu}, {McIntosh}, {Hansteen},
  {Carlsson}, {Schrijver}, {Tarbell}, {Title}, {Shine}, {Suematsu}, {Tsuneta},
  {Katsukawa}, {Ichimoto}, {Shimizu}, \& {Nagata}}]{Pontieu_2007}
{De Pontieu}, B., {McIntosh}, S., {Hansteen}, V.~H., {et~al.} 2007, \pasj, 59,
  S655

\bibitem[{{De Pontieu} {et~al.}(2011){De Pontieu}, {McIntosh}, {Carlsson},
  {Hansteen}, {Tarbell}, {Boerner}, {Martinez-Sykora}, {Schrijver}, \&
  {Title}}]{Pontieu_2011}
{De Pontieu}, B., {McIntosh}, S.~W., {Carlsson}, M., {et~al.} 2011, Science,
  331, 55

\bibitem[{{Dey} {et~al.}(2022){Dey}, {Chatterjee}, {Murthy}, {Kors{\'o}s},
  {Liu}, {Nelson}, \& {Erd{\'e}lyi}}]{Dey_22}
{Dey}, S., {Chatterjee}, P., {Murthy}, O.~V.~S.~N., {et~al.} 2022, Nature
  Physics, 18, 595

\bibitem[{{Doschek}(2012)}]{Doschek_2012}
{Doschek}, G.~A. 2012, \apj, 754, 153

\bibitem[{{Guarrasi} {et~al.}(2014){Guarrasi}, {Reale}, {Orlando}, {Mignone},
  \& {Klimchuk}}]{Guarrasi_14}
{Guarrasi}, M., {Reale}, F., {Orlando}, S., {Mignone}, A., \& {Klimchuk}, J.~A.
  2014, \aap, 564, A48

\bibitem[{{Hara} {et~al.}(2008){Hara}, {Watanabe}, {Harra}, {Culhane}, {Young},
  {Mariska}, \& {Doschek}}]{Hara_2008}
{Hara}, H., {Watanabe}, T., {Harra}, L.~K., {et~al.} 2008, \apjl, 678, L67

\bibitem[{{Judge} \& {Carlsson}(2010)}]{Judge_2010}
{Judge}, P.~G., \& {Carlsson}, M. 2010, \apj, 719, 469

\bibitem[{{Klimchuk}(2006)}]{Klimchuk_2006}
{Klimchuk}, J.~A. 2006, \solphys, 234, 41

\bibitem[{{Klimchuk}(2012)}]{Klimchuk_2012}
---. 2012, Journal of Geophysical Research (Space Physics), 117, A12102

\bibitem[{{Klimchuk}(2015)}]{Klimchuk_2015}
---. 2015, Philosophical Transactions of the Royal Society of London Series A,
  373, 20140256

\bibitem[{{Klimchuk} \& {Bradshaw}(2014)}]{Klimchuk_2014}
{Klimchuk}, J.~A., \& {Bradshaw}, S.~J. 2014, \apj, 791, 60

\bibitem[{Kosugi {et~al.}(2007)Kosugi, Matsuzaki, Sakao, Shimizu, Sone,
  Tachikawa, Hashimoto, Minesugi, Ohnishi, Yamada, Tsuneta, Hara, Ichimoto,
  Suematsu, Shimojo, Watanabe, Shimada, Davis, Hill, Owens, Title, Culhane,
  Harra, Doschek, \& Golub}]{Kosugi_2007}
Kosugi, T., Matsuzaki, K., Sakao, T., {et~al.} 2007, Solar Physics, 243, 3

\bibitem[{{Landi} {et~al.}(2013){Landi}, {Young}, {Dere}, {Del Zanna}, \&
  {Mason}}]{chianti}
{Landi}, E., {Young}, P.~R., {Dere}, K.~P., {Del Zanna}, G., \& {Mason}, H.~E.
  2013, \apj, 763, 86

\bibitem[{{Mart{\'{\i}}nez-Sykora} {et~al.}(2017){Mart{\'{\i}}nez-Sykora}, {De
  Pontieu}, {Hansteen}, {Rouppe van der Voort}, {Carlsson}, \&
  {Pereira}}]{Martinez_2017}
{Mart{\'{\i}}nez-Sykora}, J., {De Pontieu}, B., {Hansteen}, V.~H., {et~al.}
  2017, Science, 356, 1269

\bibitem[{Mart{\'{\i}}nez-Sykora {et~al.}(2018)Mart{\'{\i}}nez-Sykora, Pontieu,
  Moortel, Hansteen, \& Carlsson}]{Martinez_2018}
Mart{\'{\i}}nez-Sykora, J., Pontieu, B.~D., Moortel, I.~D., Hansteen, V.~H., \&
  Carlsson, M. 2018, The Astrophysical Journal, 860, 116

\bibitem[{{Mignone} {et~al.}(2007){Mignone}, {Bodo}, {Massaglia}, {Matsakos},
  {Tesileanu}, {Zanni}, \& {Ferrari}}]{2007ApJS..170..228M}
{Mignone}, A., {Bodo}, G., {Massaglia}, S., {et~al.} 2007, \apjs, 170, 228

\bibitem[{{Miki{\'c}} {et~al.}(2013){Miki{\'c}}, {Lionello}, {Mok}, {Linker},
  \& {Winebarger}}]{Mikic_13}
{Miki{\'c}}, Z., {Lionello}, R., {Mok}, Y., {Linker}, J.~A., \& {Winebarger},
  A.~R. 2013, \apj, 773, 94

\bibitem[{{Patsourakos} \& {Klimchuk}(2006)}]{Patsourakos_2006}
{Patsourakos}, S., \& {Klimchuk}, J.~A. 2006, \apj, 647, 1452

\bibitem[{{Patsourakos} {et~al.}(2014){Patsourakos}, {Klimchuk}, \&
  {Young}}]{Patsourakos_2014}
{Patsourakos}, S., {Klimchuk}, J.~A., \& {Young}, P.~R. 2014, \apj, 781, 58

\bibitem[{{Pereira} {et~al.}(2011){Pereira}, {De Pontieu}, \&
  {Carlsson}}]{Pereira_2011}
{Pereira}, T.~M., {De Pontieu}, B., \& {Carlsson}, M. 2011, in AGU Fall Meeting
  Abstracts, Vol. 2011, SH34B--01

\bibitem[{{Peter} \& {Judge}(1999)}]{Peter_1999}
{Peter}, H., \& {Judge}, P.~G. 1999, \apj, 522, 1148

\bibitem[{{Petralia} {et~al.}(2014){Petralia}, {Reale}, {Orlando}, \&
  {Klimchuk}}]{Petralia_2014}
{Petralia}, A., {Reale}, F., {Orlando}, S., \& {Klimchuk}, J.~A. 2014, \aap,
  567, A70

\bibitem[{{Pneuman} \& {Kopp}(1977)}]{Pneuman_1977}
{Pneuman}, G.~W., \& {Kopp}, R.~A. 1977, \aap, 55, 305

\bibitem[{{Pneuman} \& {Kopp}(1978)}]{Pneuman_1978}
---. 1978, \solphys, 57, 49

\bibitem[{{Roberts}(1945)}]{Roberts45}
{Roberts}, W.~O. 1945, \apj, 101, 136

\bibitem[{{Samanta} {et~al.}(2019){Samanta}, {Tian}, {Yurchyshyn}, {Peter},
  {Cao}, {Sterling}, {Erd{\'e}lyi}, {Ahn}, {Feng}, {Utz}, {Banerjee}, \&
  {Chen}}]{samanta_19}
{Samanta}, T., {Tian}, H., {Yurchyshyn}, V., {et~al.} 2019, Science, 366, 890

\bibitem[{Secchi(1877)}]{Secchi1877}
Secchi, P. 1877, Le soleil. 2e partie (Gauthier-Villars)

\bibitem[{{Skogsrud} {et~al.}(2015){Skogsrud}, {Rouppe van der Voort}, {De
  Pontieu}, \& {Pereira}}]{Skogsrud_2015}
{Skogsrud}, H., {Rouppe van der Voort}, L., {De Pontieu}, B., \& {Pereira},
  T.~M.~D. 2015, \apj, 806, 170

\bibitem[{{Sterling} \& {Moore}(2017)}]{Sterling_2016}
{Sterling}, A.~C., \& {Moore}, R.~L. 2017, in AGU Fall Meeting Abstracts, Vol.
  2017, SH43A--2791

\bibitem[{Tian {et~al.}(2011)Tian, McIntosh, Pontieu, Mart{\'{\i}}nez-Sykora,
  Sechler, \& Wang}]{Tian_2011}
Tian, H., McIntosh, S.~W., Pontieu, B.~D., {et~al.} 2011, The Astrophysical
  Journal, 738, 18

\bibitem[{{Tripathi} \& {Klimchuk}(2013)}]{Tripathi_2013}
{Tripathi}, D., \& {Klimchuk}, J.~A. 2013, \apj, 779, 1

\bibitem[{{Tripathi} {et~al.}(2012){Tripathi}, {Mason}, \&
  {Klimchuk}}]{Tripathi_2012}
{Tripathi}, D., {Mason}, H.~E., \& {Klimchuk}, J.~A. 2012, \apj, 753, 37

\bibitem[{{Viall} {et~al.}(2021){Viall}, {De Moortel}, {Downs}, {Klimchuk},
  {Parenti}, \& {Reale}}]{Viall_2021}
{Viall}, N.~M., {De Moortel}, I., {Downs}, C., {et~al.} 2021, in Solar Physics
  and Solar Wind, ed. N.~E. {Raouafi} \& A.~{Vourlidas}, Vol.~1, 35

\bibitem[{{Withbroe}(1983)}]{Withbroe_1983}
{Withbroe}, G.~L. 1983, \apj, 267, 825

\end{thebibliography}

\end{document}